\documentclass[a4paper,fleqn]{cas-dc}

\usepackage[authoryear]{natbib}

\usepackage{csquotes}

\def\tsc#1{\csdef{#1}{\textsc{\lowercase{#1}}\xspace}}
\tsc{WGM}
\tsc{QE}
\tsc{EP}
\tsc{PMS}
\tsc{BEC}
\tsc{DE}

\begin{document}
\let\WriteBookmarks\relax
\def\floatpagepagefraction{1}
\def\textpagefraction{.001}
\shorttitle{Exploring CoCo Challenges in ML Engineering Teams: Insights From the Semiconductor Industry}
\shortauthors{A. Azamnouri et~al.}

\title [mode = title]{Exploring Collaboration and Communication Challenges in ML Engineering Teams: Insights From the Semiconductor Industry}

\author[1]{Aidin Azamnouri}[orcid=0009-0004-8409-2316]
\cormark[1]
\ead{aidin.azamnouri@tum.de}


\affiliation[1]{organization={Chair of Software Engineering, TUM School of Computation, Information and Technology, Technical University of Munich},
                city={Heilbronn},
                country={Germany}}

\author[1]{Markus Haug}[orcid=0000-0001-9377-0677]



\author[2]{Lucas Woltmann}[orcid=0000-0003-0720-8878]

\affiliation[2]{organization={Zeiss},
                city={Oberkochen},
                country={Germany}}

\author[3]{Manuel Fritz}[orcid=0000-0003-4640-8477]        

\affiliation[3]{organization={Business School, Pforzheim University},
                city={Pforzheim},
                country={Germany}} 

\author[4]{Justus Bogner}[orcid=0000-0001-5788-0991]

\affiliation[4]{organization={Software and Sustainability (S2) Group, Vrije Universiteit Amsterdam},
                city={Amsterdam},
                country={The Netherlands}} 

\author[1]{Stefan Wagner}[orcid=0000-0002-5256-8429]

\cortext[cor1]{Corresponding author}



\begin{abstract}
The integration of machine learning (ML) into complex software systems has increased challenges in collaboration and communication (CoCo) of the teams building these systems. ML engineering (MLE) teams often involve diverse roles, ML engineers, data scientists, software engineers, and domain experts, each bringing unique goals, experiences, and jargon. These interdisciplinary dynamics can make it challenging to deploy, reproduce, and maintain ML-enabled systems over the long term. Previous studies have uncovered several CoCo challenges and practices, but most have focused on software-centric companies, leaving limited empirical understanding of how these dynamics unfold in hardware-centric contexts. In hardware-centric environments, CoCo challenges are shaped by additional constraints such as strict data governance, long development cycles, and tight coupling with physical processes, which amplify coordination complexity and reduce flexibility. To strengthen empirical understanding in such settings, we present a qualitative investigation of MLE teams within a global semiconductor company, where ML-enabled systems and manufacturing processes introduce additional complexity. We interviewed 12 practitioners regarding CoCo practices, tools, challenges, and approaches. Through analysis, we identified 16 recurring challenges, with unclear roles and responsibilities emerging as the most critical, and common practices and recommendations practitioners considered effective in mitigating CoCo problems. While grounded in a single organizational context, our findings align with known issues in interdisciplinary ML-enabled systems development, but also demonstrate how these challenges manifest differently under hardware-driven constraints. Our results highlight directions for future research and tool support to strengthen CoCo in MLE projects and ensure the success of ML-enabled systems.
\end{abstract}

\begin{keywords}
Collaboration \sep Communication \sep Machine Learning \sep ML-Enabled Systems \sep ML Engineering Team \sep Semiconductor
\end{keywords}

\maketitle

\section{Introduction}

Bringing machine learning (ML) into real-world systems is not limited to designing novel algorithms, but involves a range of activities such as data collection, data preprocessing, model training, deployment, and continuous maintenance. These activities require close interaction between practitioners from different disciplines, including ML engineers, data scientists, software engineers, and domain experts. Hence, effective collaboration and communication (referred to as CoCo in this paper) become essential to align expectations, coordinate interdependent tasks, and ensure that ML-enabled systems can be successfully developed, deployed, and maintained over time~\citep{Krause_J_ttler_2022, Nahar_2022, Mailach_2023}.
Prior studies have explored challenges in software engineering for ML systems and machine learning operations (MLOps) in general~\citep{Amershi_2019, Lima_2022, Honkanen_2022, Haug_2025}, and highlight the complexities of integrating ML into software systems~\citep{Wan_2020, Kalinowski_2025}. This emphasizes the need for effective interdisciplinary collaboration among data scientists, ML engineers, and software developers~\citep{Nahar_2022, Mailach_2023, Busquim_2024, Azamnouri_2025}.

While ML engineering (MLE) in software-centric companies already requires the collaboration of many different roles~\citep{Amershi_2019, Busquim_2024}, this becomes even more problematic in the semiconductor industry. Data scientists, domain experts, software engineers, sensor and imaging specialists, optics engineers, physicists, and hardware specialists, each with differing assumptions, vocabularies, and priorities, need to collaborate effectively. Misalignment in team understanding about goals, data ownership, or model deployment processes can introduce friction, increase technical debt~\citep{Sculley_2015, Polyzotis_2018, Recupito_2024}, and impede the effective translation of ML prototypes into production systems~\citep{Zaharia_2018, Lewis_2021, Eken_2025}. Such issues are compounded by the rapid pace of innovation and the critical demand for high reliability in semiconductor production pipelines~\citep{Li_2023, Xu_2024}.

Recent research has underscored the growing importance of socio-technical factors in ML deployment~\citep{Amershi_2019, Sambasivan_2021, Mailach_2023}, highlighting how CoCo barriers, such as unclear team roles, mismatched expectations, and inadequate feedback loops, can hinder model adoption and maintenance. However, most of these studies focus on software companies. The semiconductor domain, characterized by proprietary data, stringent precision requirements, long development cycles, and very rigid deadlines, presents unique CoCo challenges that remain underexplored.
Conducting a study within a global semiconductor company is therefore important for understanding CoCo challenges in high-tech manufacturing industries.

Moreover, while previous work, such as~\citet{Nahar_2022} or~\citet{Busquim_2024}, explored CoCo challenges and practices of MLE teams in the software-centric industry, very few studies have deeply analyzed CoCo challenges across various MLE team members of a \textit{single} hardware-centric company. Hence, our study aims to contribute a stronger empirical understanding of CoCo phenomena in contexts where ML integration intersects with high-stakes, precision-driven production processes.

To start closing this gap, we studied the CoCo of the MLE teams of a global hardware-focused company via interviews. This company is a leader in semiconductor manufacturing technologies. It operates in a highly complex, multidisciplinary domain where teams of software engineers, data scientists, optical physicists, mechanical engineers, and systems integrators must work closely together. This environment exemplifies the intersection of advanced engineering disciplines with strict performance and quality demands, making effective collaboration essential yet difficult to achieve.

In this company, there may be weeks or months of production time for a single product due to the very high requirements regarding the accuracy of the production. Training ML models in this context poses additional challenges in terms of sample size. For instance, the company generates large volumes of imaging data that are processed through a modular ML workflow designed for defect detection and classification in customer products. The organization adopts both data-driven and process-driven manufacturing approaches to continuously optimize production processes, thereby improving overall product quality and efficiency. Fulfilling this commitment demands highly customized solutions, which are developed in-house through close collaboration among data scientists, software engineers, and physicists.

In the studied hardware-centric environment, ML projects are typically embedded within larger software–hardware systems rather than developed as standalone components. These projects follow a socio-technical structure in which ML solutions are tightly coupled with physical manufacturing processes, sensor systems, and production software. As a result, project organization, workflows, and system architectures differ substantially from those commonly reported in software-only ML settings.

The workflow usually starts from a production or manufacturing problem, e.g., defect detection, process optimization, or quality inspection, rather than from a purely data-driven question. Domain experts define constraints and performance requirements grounded in physical processes and hardware limitations. Data collection and labeling are strongly influenced by sensor availability, measurement precision, and data governance constraints. ML model development is conducted iteratively, often with small or highly imbalanced datasets, requiring close feedback loops between ML practitioners and domain experts to validate assumptions and intermediate results. Integration into production systems involves software engineers embedding trained models into existing pipelines, ensuring compatibility with hardware interfaces, latency requirements, and reliability standards. Deployment is therefore not a single handover step but a negotiated process that spans development, validation, and gradual rollout.

Architecturally, ML components are typically embedded as modular services within larger software systems that interface directly with hardware. A common pattern is a layered architecture consisting of data acquisition layers connected to sensors and machines, data preprocessing and feature extraction pipelines, ML model components for inference or decision support, and downstream software systems that consume ML outputs for visualization, control, or optimization. These components are connected through APIs, shared data repositories, or message-based interfaces, and must comply with strict performance, traceability, and documentation requirements. Importantly, architectural decisions are often constrained by legacy systems and regulatory or organizational policies, further complicating collaboration across roles.

By explicitly situating ML solutions within these integrated software–hardware project structures, our study highlights why CoCo challenges in such environments are more pronounced than in software-centric contexts. The need to align physical constraints, ML workflows, and software architectures makes CoCo a critical factor for the successful development and deployment of ML-enabled systems in the semiconductor industry.

Investigating such a setting provides valuable insights into the real-world CoCo issues that arise in safety-relevant and innovation-driven software-intensive systems. Furthermore, lessons learned from this company that operates in approximately 50 countries might be generalized to companies in similar industrial sectors. This study, therefore, contributes to both academic understanding and practical strategies for improving CoCo factors in industry. It is important to note that insights from software-centric ML engineering are not invalid in hardware-focused environments. Many previously identified CoCo challenges are also observed in our study and appear to be fundamental to the socio-technical nature of ML-enabled systems. However, rather than emerging in identical forms, our findings suggest that these challenges are significantly shaped by the surrounding industrial context. Consequently, this study does not position hardware-centric MLE as entirely distinct from software-centric settings, but instead as an extension where known challenges are recontextualized and, in some cases, amplified by domain-specific conditions.

In this paper, we describe a qualitative study of MLE teams in a leading semiconductor company to identify and analyze the CoCo challenges that affect their workflow. Through interviews and analysis, we offer insights into how teams manage collaboration between diverse roles and communication across organizational layers. Our findings aim to inform both practitioners seeking to improve cross-functional ML practices and researchers interested in the socio-technical dynamics of applied ML in industrial contexts.

In this paper, we provide the following contributions:
\begin{itemize}
    \item Providing one of the first in-depth qualitative analyses of CoCo in ML engineering within a global semiconductor company, highlighting how hardware constraints, data governance, and manufacturing processes shape ML-enabled system development.
    \item Identifying 16 CoCo challenges, distinguishing between widely known issues in ML-enabled systems and those that are contextually shaped by the semiconductor domain and matrix organizational structure.
    \item Reporting concrete practices and coordination mechanisms used by practitioners to address CoCo challenges, revealing both effective bottom-up strategies and gaps where systematic organizational support and future research are needed.
\end{itemize}

The remainder of this paper is structured as follows: Section 2 reviews related work. Section 3 details the study design. Section 4 presents the results, followed by a discussion in Section 5. Section 6 addresses the main threats to validity and limitations of the study. Finally, Section 7 concludes the paper and outlines directions for future research.

\section{Related Work}

The development of ML-enabled systems has introduced unique CoCo challenges, arising from the interdisciplinary nature of integrating software engineering and ML workflows. In this regard,~\citet{Nahar_2022} conducted a large-scale qualitative study based on 45 interviews across 28 organizations and identified four central categories of collaboration challenges: communication, documentation, engineering, and process. Their findings show that communication becomes particularly problematic due to cultural and epistemic differences between roles, especially as data scientists and ML engineers often operate with different assumptions, priorities, and terminologies. For example, model requirements are frequently underspecified or evolve over time, requiring data scientists to continuously educate stakeholders about the realistic capabilities and limitations of ML systems.

Beyond communication,~\citet{Nahar_2022} also highlight documentation challenges, where traditional software documentation practices are often insufficient for capturing data dependencies, model behavior, and experimental workflows. Engineering challenges emerge from the need to integrate data-centric and code-centric processes, while process-related challenges stem from the lack of standardized development lifecycles for ML-enabled systems. These findings collectively emphasize that CoCo breakdowns are not isolated issues but are deeply embedded in the socio-technical structure of ML development.

While this work provides a comprehensive view of CoCo challenges in software-centric ML environments, it primarily reflects organizations where ML systems are developed as digital products or services. In contrast, our study examines how these well-documented challenges manifest in a hardware-centric, manufacturing-driven context. As we show later, challenges such as communication gaps and unclear requirements persist, but are further shaped by domain-specific constraints such as strict data governance, long production cycles, and tightly coupled physical processes. This suggests that existing taxonomies, such as the one proposed by~\citet{Nahar_2022}, provide a useful foundation, but require contextualization to fully capture CoCo dynamics in industrial ML settings.

Moreover,~\citet{Piorkowski_2021} investigated the everyday realities of AI developers, particularly focusing on how data scientists collaborate with colleagues from diverse disciplinary backgrounds. Through interviews and an analysis of communication artifacts, the study revealed a recurring knowledge mismatch, where experienced data scientists must effectively translate their work into forms that are understandable to non-ML stakeholders. This translation goes beyond simplifying terminology; it involves reframing probabilistic outputs, uncertainty, and model limitations into actionable insights that align with business or operational goals. The authors further show that this translation work is continuous and situational, often occurring during key collaboration moments such as requirement elicitation, result interpretation, and decision-making. As a result, data scientists frequently take on the role of intermediaries, bridging gaps between technical and non-technical perspectives. This additional responsibility can increase cognitive load and shift effort away from core development tasks, highlighting that communication challenges in ML-enabled systems are not merely incidental but integral to the work itself.

While~\citet{Piorkowski_2021} primarily examine knowledge translation in software-oriented AI teams, our study extends this perspective by considering how such translation challenges manifest in a hardware-centric, manufacturing environment. In this context, the need for translation is further intensified by domain-specific factors, such as physics-driven processes, specialized terminology, and strict production constraints. Consequently, aligning understanding across roles requires not only bridging ML and software perspectives but also integrating domain expertise from fields such as engineering and applied physics, making CoCo even more demanding.

\citet{Busquim_2024} explored the collaboration dynamics between software engineers and data scientists in ML-enabled systems, with a particular focus on soft skill–related challenges. Their findings highlight a fundamental communication gap caused by the lack of a shared vocabulary and differing mental models. While software engineers tend to prioritize system architecture, scalability, and deployment concerns, data scientists focus more on model development, experimentation, and data analysis. This divergence often leads to misaligned expectations, misunderstandings about requirements, and friction during integration phases, especially when transitioning models from prototyping to production environments. Beyond identifying these gaps, the study also emphasizes that successful collaboration requires more than technical alignment; it depends on the ability of team members to negotiate meaning, align priorities, and develop shared representations of system goals. In this sense, communication challenges are closely intertwined with coordination practices and organizational structures, reinforcing the socio-technical nature of ML-enabled system development.

Building on this work,~\citet{Busquim_2024_2} further investigated how collaboration between software engineers and data scientists can be improved in practice. Through two focus group sessions with industry practitioners, they identified concrete tasks where close collaboration is beneficial and essential, such as defining data access rules, specifying interfaces between ML components and software systems, and deploying trained models into production pipelines. These tasks represent critical coordination points where misalignment can significantly impact system performance and project outcomes.

While these studies provide important insights into collaboration challenges and practices in software-centric ML development, they primarily focus on interactions between software engineers and data scientists. In contrast, our study considers a broader set of roles, including domain experts such as process engineers and physicists, and examines how collaboration unfolds in a hardware-centric, manufacturing environment. In such contexts, the communication gaps identified by~\citet{Busquim_2024} are present and are further amplified by additional domain-specific constraints and the need to align ML outputs with physical production processes. This highlights the need to extend existing findings to more complex interdisciplinary settings.

Furthermore,~\citet{Almahmoud_2021} examined how teams communicate about the quality of ML models within a large international technology company. Through interviews with 15 practitioners, followed by a broader survey of 168 team members, the authors demonstrate that the notion of quality in ML-enabled systems extends far beyond traditional technical metrics such as accuracy, precision, or overfitting. Instead, practitioners adopt a more holistic understanding that includes ethical considerations, legal compliance, operational readiness, and user satisfaction, reflecting the inherently socio-technical nature of ML systems. The study highlights that this broadened definition of quality introduces significant communication challenges. Different stakeholders interpret quality through the lens of their own expertise. Technical staff often rely on ML-centric metrics and probabilistic reasoning, whereas product managers and UX professionals focus on user impact, usability, and trust. This divergence leads to difficulties in aligning expectations, evaluating system performance, and making informed decisions about deployment.

While this work provides important insights into how quality-related communication unfolds in software-centric ML environments, our study extends this perspective by examining how such multi-dimensional notions of ML and technical terms are discussed in a hardware-centric, manufacturing context. Consequently, communication about ML model quality and capabilities must not only bridge technical and user-oriented perspectives but also account for the potential impact on costly and irreversible manufacturing processes, thereby increasing both the complexity and the criticality of CoCo.

Taking a slightly different perspective,~\citet{Wu_2024} revisit the V-Model, a well-established systems engineering methodology, to explore how it can support the development of ML-enabled software systems. Through interviews with 11 professionals from mid-size and large technology firms, including both software developers and ML practitioners, the authors investigate real-world collaboration challenges that arise as ML models transition from prototyping to production environments. Their analysis results in eight practical propositions that illustrate how the structured nature of the V-Model can help address common CoCo issues in ML projects. 

In particular,~\citet{Wu_2024} highlight that the V-Model's emphasis on clear phase separation, traceability, and validation can mitigate challenges such as responsibility confusion, fragmented documentation, and mismatched understanding across teams. By explicitly linking development stages with corresponding validation activities, the model encourages teams to define requirements more clearly and to establish shared expectations early in the process. This structured approach contrasts with the often exploratory, iterative nature of ML development, suggesting that integrating systems engineering principles can provide much-needed scaffolding for coordination. While~\citet{Wu_2024} focus on software-centric organizations, their insights are particularly relevant for our study, as the semiconductor domain has strong roots in systems engineering practices. However, in our context, the challenges addressed by the V-Model, such as responsibility clarity, documentation, and alignment, are further intensified by hardware constraints, long production cycles, and strict validation requirements.

In addition, several studies have examined CoCo challenges in ML-enabled systems from a community smells and socio-technical perspective, shifting the focus from individual interactions to systemic organizational patterns. For example,~\citet{Annunziata_2025} identified recurring community smells in ML-enabled systems, such as siloed expertise, role misunderstandings, and missing engineering competence. These issues capture persistent coordination inefficiencies that arise when teams lack shared practices, clear ownership structures, or sufficient cross-functional integration. The authors further link these smells to their underlying causes, such as organizational fragmentation and uneven skill distribution, and discuss their potential effects on productivity, knowledge sharing, and system quality.

Similarly,~\citet{Mailach_2023} analyzed socio-technical anti-patterns in ML-enabled software development by studying public talks from industry leaders. Their findings highlight recurring organizational issues, including fragmented responsibility, communication breakdowns, and the separation of data, model, and engineering concerns across teams. These anti-patterns emphasize that many CoCo challenges are not merely operational but are deeply rooted in organizational design and decision-making processes, often requiring structural interventions rather than purely technical solutions.

Together, these studies provide higher-level abstractions of CoCo challenges, framing them as systemic socio-technical phenomena rather than isolated team-level issues. However, they primarily focus on software-centric environments or leadership perspectives, where ML systems are developed as digital products and where organizational dynamics may differ significantly from those in industrial settings. Our study complements and extends this line of work by grounding these abstract concepts in a hardware-centric semiconductor context. In this environment, issues such as siloed expertise and fragmented responsibility are not only present but are further intensified by domain-specific constraints. Moreover, while prior work often identifies these issues at a conceptual level, our findings provide detailed empirical accounts of how such challenges manifest in day-to-day practices, thereby offering a more fine-grained understanding of CoCo dynamics in complex industrial ML systems.

In summary, while prior research on CoCo in ML-enabled systems has primarily focused on stable, software-centric environments, our interview study addresses a critical gap by examining these challenges in the high-variability, hardware-centric context of the semiconductor industry. In this domain, each product is unique, necessitating meticulous coordination and cross-functional awareness to ensure alignment across teams and tools. Our study highlights the need for robust CoCo mechanisms, offering novel insights into the intersection of ML and software engineering practices and high-tech hardware development.

\begin{figure*}[ht]
\centerline{\includegraphics[width=1\textwidth]{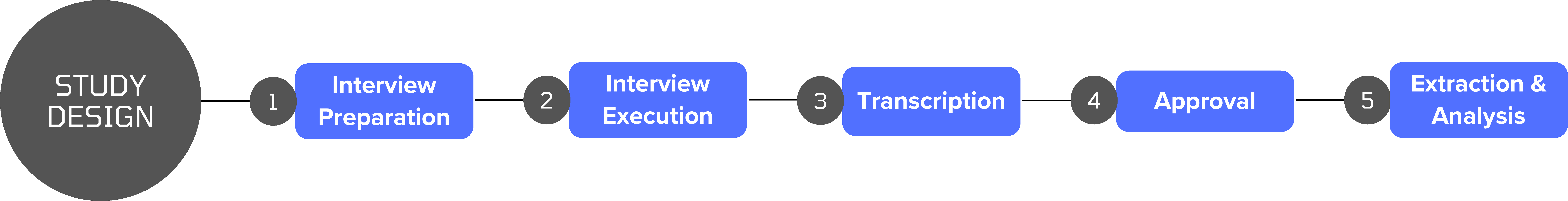}}
\caption{Overview of the overall research process.}
\label{fig1}
\end{figure*}

\section{Study Design}

We employed a qualitative approach via semi-structured interviews~\citep{Hove_2005, Adams_2015}. This allows for flexibility in exploring participants' insights while maintaining consistency across interviews. The study was composed of five key phases that are illustrated in Fig.~\ref{fig1} and detailed in each subsection below.
Our primary objectives were to understand MLE team members' roles and responsibilities, practice and tool choices, the CoCo challenges they faced, and their suggestions for addressing challenges. To achieve the goals of this study, we articulated the following research questions:

\textbf{RQ1}: What are the usual collaboration and communication practices and tools MLE teams use in the hardware-centric company?

\textbf{RQ2}: What are the collaboration and communication challenges faced by MLE teams in the hardware-centric company?

\textbf{RQ3}: How can practitioners deal with collaboration and communication challenges in the hardware-centric company?

\subsection{Interview Preparation}

We began by developing a semi-structured interview protocol using~\citet{Hove_2005, Runeson_2008} to ensure both consistency across sessions and flexibility to follow up on relevant emergent themes. The protocol was iteratively refined through two pilot interviews with software practitioners familiar with ML development but not participating in the main study. These pilots helped ensure clarity, flow, and appropriateness of the questions. Additionally, we prepared a study preamble, a privacy statement, and a consent form. The study preamble provides an overview of the research objectives and methodology. The privacy statement is used to inform participants about how their personal data will be collected, used, and protected throughout the study. The consent form is designed to inform participants about the study and obtain their voluntary agreement to participate. Prior to participant recruitment, the employee representative committee (workers' council) examined the study protocol and related materials to confirm they met all ethical and legal standards. After that, interviewees received the study overview, privacy information, and consent form in advance of the interview and were asked to review and return the signed consent form beforehand.

The final protocol included questions focusing on participants' demographics (including roles and experience with ML systems), the CoCo challenges they faced, the practices and tools they use, and their suggestions for addressing CoCo issues.
The full protocol with the detailed questions and all other shareable study artifacts can be found in our supplementary materials.\footnote{\url{https://doi.org/10.5281/zenodo.17358331}}
Participant recruitment was conducted via purposive and snowball sampling, targeting professionals with experience in building, deploying, or maintaining ML-enabled systems within the company. All participants were provided with detailed information sheets and consent forms in compliance with the research ethics guidelines.

\subsection{Interview Execution}

We conducted 12 interviews, each lasting approximately 30 to 45 minutes. Interviews were conducted remotely via a secure video conferencing platform. All sessions were audio-recorded with the explicit consent of the participants. Interviews were led by the first two authors, one serving as the primary facilitator and the other taking observational notes to capture non-verbal cues and potential follow-up areas. To ensure participants fully understood the consent form, we requested verbal confirmation for audio recording in addition to the signed consent. Once verbal consent was given, we proceeded to record the interview audio for transcription purposes.

We initially intended to conduct individual interviews to prevent mutual influence among participants and to allow sufficient time for in-depth discussion. However, in two instances, participants preferred to be interviewed in pairs, and we respected their choice. At the start of each interview, a company representative briefly reintroduced the study to participants, emphasized the workers' council's approval, and explained the legal agreements of the study. Once this introduction was completed, the representative left the call, and the interview started.

We followed a semi-structured guide primarily composed of open-ended questions. This approach allowed us to gather rich, detailed insights while maintaining flexibility to explore topics in depth.
The interview guide was divided into two sections, with the first section asking participants to evaluate their experience in areas such as software engineering (SE), software operations (SE Ops), ML, development and operations (DevOps), and MLOps.

In the second part, we inquired about the MLE team members' primary responsibilities, how they collaborate and communicate with each other, which communication channels they use, how they document essential decisions, the most significant challenges they faced while collaborating and communicating with different roles within their teams, and what strategies and approaches they leverage or suggest to cope with the challenges.
At the end of the interview, we invited participants to share any additional thoughts they felt had not been addressed to gather potential insights that might have been overlooked.

\subsection{Transcription}

After conducting the interviews, the audio recordings were transcribed using an automatic transcription tool called aTrain. With this open-source tool, any audio or video format can be transcribed and analyzed using OpenAI's Whisper transcription model~\citep{Haberl_2024}. This significantly reduced effort while increasing accuracy by minimizing human errors. As several participants requested to be interviewed not in English, we utilized aTrain's automated translation feature to accommodate this. Importantly, no interview data was shared with external entities during this process, ensuring that confidentiality and data protection were maintained.

Following that, the transcripts were manually reviewed by the authors using oTranscribe\footnote{\url{https://otranscribe.com/}} to correct obvious transcription errors and to annotate inaudible or ambiguous parts. Our final step was to export each transcript into a Microsoft Word file so that it could be approved and analyzed by participants.

\subsection{Approval}

After reviewing the transcripts for each interview, we shared them with the respective interviewees for their review and approval, ensuring that participants' voices and perspectives were accurately respected. We also requested clarification on specific terms we had marked for the transcription. Participants were free to revise or remove any parts of their statements. To track these changes, we used Microsoft Word's change tracking feature, allowing us to clearly see the edits made by participants. Once the interviewees submitted their revisions, we carefully incorporated their changes into the final transcripts. When a transcript was approved, we permanently deleted the corresponding audio recording and relied exclusively on the transcripts for all further analysis.

\subsection{Extraction and Analysis}

In the last step, we analyzed the prepared transcripts using qualitative analysis to identify patterns and themes across the data~\citep{Seaman_1999}. First, we selected those parts of the interview transcripts that were most relevant to the research questions, focusing on descriptions of CoCo practices, challenges, and mitigation strategies. These segments formed the basis for the subsequent coding process. Then, the first two authors independently conducted an initial round of open coding~\citep{Cruzes_2011} on a subset of the transcripts. During this phase, each coder assigned descriptive labels to meaningful units of text, e.g., sentences or short passages, without relying on predefined categories. This resulted in an initial set of fine-grained codes capturing observed phenomena such as role ambiguity, communication breakdowns, or documentation issues.

Following the independent coding phase, the two authors conducted a joint calibration session in which they systematically compared their codes. During this session, overlapping codes were merged, unclear labels were clarified, and differences in interpretation were discussed in detail. Disagreements were resolved through consensus, and code definitions were refined to ensure a shared understanding. The outcome of this step was a preliminary codebook containing code names, definitions, and representative examples.

This process was iteratively repeated. Additional transcripts were coded using the evolving codebook, followed by further discussion sessions to refine code definitions, merge redundant codes, and ensure consistent application. Iterations continued until no substantial ambiguities remained and the codebook stabilized. Once a stable codebook was established, it was applied to the full dataset. During this phase, previously coded transcripts were revisited where necessary to ensure consistency with the final code definitions. Given the exploratory and interpretive nature of this study, we did not compute a formal inter-coder reliability statistic. Instead, reliability was ensured through repeated joint calibration, iterative refinement of the codebook, and consensus-based resolution of disagreements, which are widely used strategies in qualitative research to ensure analytical rigor~\citep{Guest_2012, Saldana_2021}.

After coding was completed, we grouped related codes into higher-level themes representing recurring CoCo challenges and practices. For each theme, we developed a clear definition and recorded the number of interviews in which it appeared. These themes form the basis of the findings presented in the next section. Due to confidentiality constraints, full interview transcripts and extractions cannot be shared. We are not permitted to distribute or disclose this information. However, to ensure transparency, we provide representative anonymized excerpts throughout the paper to illustrate how codes and themes are grounded in participant statements. In addition, we report recurring patterns across multiple interviews rather than isolated observations, reducing the influence of individual perspectives. The analysis was conducted by the first and second authors, while the remaining authors reviewed and validated the resulting themes.

\begin{table*}[ht]
\caption{Overview of interview participant roles and experiences (in years)}
\label{tab1}
\begin{tabular}{llrrrrrr}
\toprule
\textbf{Participant ID} & \textbf{Role} & \textbf{In Role} & \textbf{SE} & \textbf{SE Ops} & \textbf{ML} & \textbf{DevOps} & \textbf{MLOps} \\
\midrule
P01  & Data Scientist          & 1     & 0     & 0     & 3.5   & 0     & 0 \\
P02  & Software Developer      & 4     & 25    & 0     & 8     & 4     & 1 \\
P03  & Software Developer      & 1     & 2     & 0     & 2     & 2     & 1 \\
P04  & System Developer        & 2.5   & 5     & 0     & 5.5   & 0     & 0.5 \\
P05  & Data Scientist          & 3     & 4     & 0     & 3     & 0     & 0 \\
P06  & Data Scientist          & 0.5   & 10    & 0     & 5     & 0     & 0 \\
P07  & Process Engineer        & 5     & 0     & 0     & 5     & 0     & 0 \\
P08  & ML Engineer             & 2     & 12    & 1.5   & 6     & 0     & 0 \\
P09  & Data Scientist          & 6     & 10    & 0     & 13    & 2     & 0 \\
P10 & Data Scientist          & 3     & 10     & 0     & 3     & 2     & 0 \\
P11 & Data Scientist          & 2.5   & 7.5   & 1.5    & 7.5   & 1     & 1 \\
P12 & Physicist               & 2     & 9     & 2      & 7     & 2     & 0 \\
\bottomrule
\end{tabular}
\end{table*}

\subsection{Participants}

We interviewed 12 professionals in 10 interviews, each bringing a unique blend of technical background and role, which is depicted in Table~\ref{tab1}. Their experiences spanned a range of industry positions, including data scientists, software and systems developers, a process engineer, an ML engineer, and a physicist.
The participants varied significantly in their primary roles and domain expertise. For example, among the five data scientists, one was relatively new to the role with only five months of experience, but brought over a decade of software engineering experience and five years in ML. We also interviewed a process engineer who had no formal software engineering or DevOps background but reported five years of experience applying ML in their domain. This perspective added depth to our understanding of how ML practices are adopted beyond traditional engineering roles.

Overall, interviewees brought between 2 and 25 years of software engineering experience and 1 to 13 years in ML. Their involvement in DevOps and MLOps ranged from none to several years, highlighting the diversity of workflows, role expectations, and skill integrations encountered across this organization. This diversity was crucial for capturing a broad and realistic picture of how ML-enabled systems are developed, deployed, and maintained in practice.

\section{Results}

We now present the findings from our interview study, which reveal key insights into current practices, challenges, and solution approaches regarding CoCo. The results are organized around the major themes that emerged during our analysis, supported by representative opinions to illustrate participants' perspectives.

\subsection{Roles and Responsibilities}

Building ML-enabled systems often requires close collaboration among team members with diverse skill sets. Based on the participant accounts, we outline below the key roles and responsibilities that they identified as central to the work of MLE team members. These insights reflect how interviewees perceived actual day-to-day responsibilities, rather than a prescriptive or formalized model. However, the company follows a matrix organization, meaning there is no centralized MLE team; instead, individuals are engaged in projects dynamically, depending on their expertise and the specific needs of each initiative. Notably, several roles overlap in practice, which, according to participants, has a significant impact on CoCo within projects. The following role descriptions capture both the intended distinctions and the practical ambiguity observed in the organization.

\begin{itemize}
    \item \textit{Data scientists / ML engineers}: Managing all aspects of the ML lifecycle, data extraction, data engineering, data preparation, data analysis, data sources integration, developing and using stochastic models, model training, model evaluation, visibility studies, ML model engineering, ML pipelines, collecting performance requirements, communicating ML concepts to non-experts, system selection and problem-solving decision, working with cloud and log, process optimization, and handling operations tasks.
    \item \textit{Software engineers}: Setup, maintain, and manage software infrastructure, define, maintain, update, upgrade, and add features to the pipelines, ML integration in software, building and configuring cloud infrastructure, ML systems production, producing data, algorithm integration in machine workflows, and moving results data to databases.
    \item \textit{System developers}: Maintain and manage software infrastructure, integration, define, maintain, update, upgrade, and add features to the pipelines, tracking, logging, and working with ML pipelines.
    \item \textit{Process engineers}: Working with the labeling tool, designing and ensuring ergonomics, developing concepts and ideas.
    \item \textit{Physicists}: Algorithm development, enhancing general algorithm capabilities, identifying optimal algorithms, and model training.
    \item \textit{Algorithm developers}: Develop the ML solutions (act as ML engineers).
    \item \textit{Development engineers}: Reliability assessment, failure mode and effects analysis, and optimizing production processes.
\end{itemize}

This wide distribution of responsibilities across roles reveals significant overlap and ambiguity, particularly in tasks related to data handling, ML pipeline development, and model deployment. In particular, algorithm developers and physicists are differentiated by their emphasis on algorithmic design vs. physical process knowledge, while development engineers and process engineers differ in their focus on system reliability and production robustness vs. operational workflows and tooling.
Participant P04 mentioned that the responsibilities typically associated with ML engineers are often assumed by algorithm developers who design and implement ML solutions. P01 stated that data scientists are primarily responsible for executing the entire ML pipeline, encompassing data extraction, data engineering, and model development. This broad scope of responsibility stems from a lack of more specialized roles within most teams, such as dedicated data engineers and ML engineers. As a result, despite their title, data scientists often function as generalists, managing end-to-end workflows due to the limited availability of domain-specific expertise across the organization.

While the mentioned roles are conceptually distinct, participants emphasized that organizational constraints often result in individuals simultaneously spanning multiple roles. For example, algorithm developers and physicists may both engage in model development and training, whereas development and process engineers may jointly shape ML requirements through their involvement in production optimization and quality assurance.
Consequently, role boundaries are frequently blurred in practice, which directly contributes to the CoCo challenges discussed in our findings.

\subsection{CoCo Practices and Tools}

To address \textbf{RQ1}, the use of various CoCo practices and tools by the MLE teams in this company was investigated. Both in-person and virtual meetings are held regularly among various roles, including domain experts, internal stakeholders, and data scientists. In-person meetings typically occur for internal discussions at the office or when team members from different locations come together. However, given the company's distributed structure, with offices located across multiple countries and cities, most collaboration takes place remotely via Microsoft Teams. Email communication, primarily through Outlook, is also a common means of exchanging information. The teams extensively utilize Azure DevOps for collaboration, leveraging its functionalities for version control, continuous integration, and project tracking. Agile practices are followed throughout the development process to enhance coordination and adaptability among team members.

When projects involve multiple contributors working on different components, a mediator often facilitates communication, ensuring a structured and efficient information flow. Within Microsoft Teams channels, software architects oversee project assignments to developers, define project statuses, and maintain related documentation. Additionally, a dedicated ML-focused group, referred to as the \textit{Community of Practice}, provides guidance and support when team members encounter technical challenges or require expert insights. The organization follows a matrix structure, where superiors have limited direct involvement in domain-specific topics. Instead, communication mainly flows through product owners and project managers, which helps maintain coherence and alignment across teams. In cases where collaboration depends primarily on data sharing, team members use Microsoft SharePoint or a network drive, e.g., Samba, as the preferred platform. Finally, certain inter-team relationships function in a customer–supplier manner, where one team specifies the requirements and expected outcomes, while the other is responsible for implementation.

Moreover, documenting each step in the development of ML-enabled systems is critical to ensuring transparency, reproducibility, and maintainability~\cite{Latendresse_2024, Indykov_2025, Assres_2025}. Within the industrial organization studied, documentation practices vary across projects, reflecting the diverse needs and preferences of practitioners. System-level documentation is often captured in PowerPoint presentations, while (RESTful) APIs are documented using OpenAPI specifications. Practitioners document datasets used for training, including their structure and contents, and maintain records of experiments, model versions, and user interfaces. Furthermore, projects are documented in README.md files, which typically include installation instructions and usage guidelines.

Architecture Decision Records (ADRs) and data contracts are also common formats, particularly for capturing design rationale and data dependencies. Besides, there is an explicit reporting process with milestones and required documentation that is sometimes written in natural language. Documentation is also co-located with code in the system's repository and is shared manually across the project team. Additional artifacts are stored in wiki pages, often used to record meeting discussions. There also exist ticketing systems as a means of documentation. In some cases, the training data itself serves as the implicit specification of the ML component. While there is an explicit documentation process aligned with project milestones, individual practices vary; some practitioners keep private notes, and a few believe their systems require minimal documentation.

These findings illustrate that CoCo in MLE projects is not tied to a single platform or practice, but emerges from a layered ecosystem of tools, processes, and organizational arrangements that reflect both the complexity of ML-enabled systems and the matrix structure of the company.

\subsection{CoCo Challenges}

\begin{figure*}[ht]
\centerline{\includegraphics[width=0.8\textwidth]{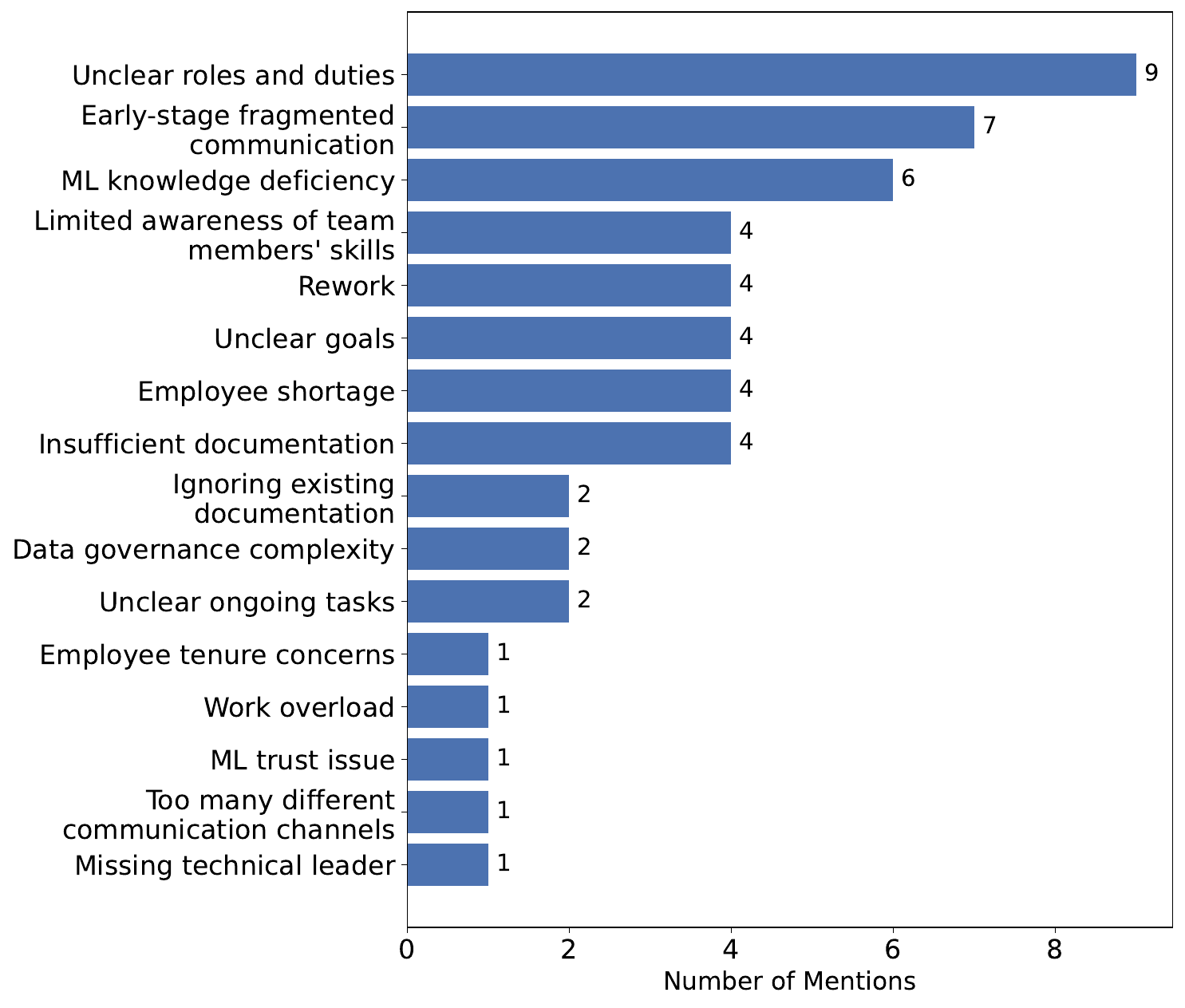}}
\caption{CoCo Challenges identified in the interviews with the number of mentions}
\label{fig2}
\end{figure*}

To answer \textbf{RQ2}, we examined the CoCo challenges practitioners at the company faced while building ML-enabled systems, with novel challenges not identified in prior work explicitly highlighted in the text using bold for clarity. As shown in Fig.~\ref{fig2}, 16 challenges were identified.

The most mentioned challenge is \textit{Unclear roles and duties} (C1). Without a well-defined structure, team members may struggle to understand who is responsible for tasks such as data collection, model development, deployment, and monitoring. For example, a data scientist might be responsible for all operations in a department. In a large enterprise, responsibilities are vast, and finding the right person for specific information is often challenging. P11 said that responsibilities for areas such as data modeling, data storage, and data platform infrastructure are often assumed to lie with other departments, but role boundaries are not always clearly defined. For instance, while the R\&D team focuses primarily on data cleaning processes and relies on existing MLOps platforms, it is not responsible for providing or maintaining these underlying solutions. Similarly, P01 mentioned that there are different groups at the company without clear orders of who is responsible for what.
This challenge has also been widely reported in software-centric ML contexts; however, in our setting, it is amplified by the matrix organization, where responsibilities are distributed across departments with limited visibility, making role boundaries more difficult to establish and maintain.

\textit{Early-stage fragmented communication} (C2) was the other challenge mentioned by 7 participants. People working on the same topic without effective communication, or those who are not focusing on an important topic due to a lack of enough communication, especially during the initial steps, may lead to inefficiencies, duplicated efforts, and missed opportunities for innovation. In P08's opinion, misunderstandings about the capabilities and implementation requirements of large language models (LLMs) can further exacerbate early-stage communication challenges. For instance, some stakeholders may assume that tasks such as automatically generating outputs, e.g., X, Y, and Z, can be implemented and will lead to significant time savings. However, upon closer examination, these assumptions often overlook the additional time, effort, and resources required for integration, customization, and ongoing operation. Such discrepancies highlight a broader lack of shared understanding regarding what LLMs can realistically achieve. While early-stage communication issues are known in prior work, they manifest differently here due to long development cycles and tight coupling with manufacturing processes, where early misunderstandings can propagate and become costly to correct later.

Six participants identified \textit{ML knowledge deficiency} (C3) as a challenge. Team members who lack ML knowledge might have some unrealistic expectations, leading to misunderstandings about the time and resources required to develop and deploy ML-enabled systems. Internal and external stakeholders are not aware that developing ML solutions has its own complexities and associated costs that do not exist for traditional software~\citep{Bhatt_2020, Suresh_2021}. P02 remarked that a common challenge arises when a production issue exists that could be addressed using ML, but remains unresolved since they are unaware that ML could offer a viable solution. In addition, P04 commented that there is a common misconception that ML components can be developed and integrated just like any other software feature. Similarly, P05 frequently deals with colleagues from the production department who have limited familiarity with ML, so they must find ways to explain complex ML concepts in simpler terms. Similar knowledge gaps have been observed in software-centric settings, but in this context, they extend beyond software concerns to include the interaction between ML models and precision-critical physical processes, increasing the complexity of communication.

The concern about \textit{\textbf{Limited awareness of team members' skills}} (C4) was highlighted by 4 respondents. This would lead to inefficient task delegation and missed opportunities for collaboration. P05 said that \enquote{The biggest problem is that maybe team members do not know what I know, and I do not know what they know.}
Unlike general role awareness issues reported in prior work, this challenge is intensified by the matrix structure, where dynamically formed teams and distributed expertise reduce visibility into available skills across departments.

According to 4 participants, \textit{Rework} (C5) was also a significant obstacle. Building ML systems repeatedly and explaining what ML is to others multiple times from scratch would be time-consuming and cause redundant efforts and confusion among team members. P09 indicated that rework is closely tied to persistent challenges in achieving a shared understanding of ML requirements and processes. This participant highlighted the difficulty of aligning stakeholders on both needs and objectives, noting that substantial effort is often required to reduce discussions to a well-defined task with a common knowledge base. Despite these efforts, iterative processes tend to progress slowly, as the same fundamental concepts must be repeatedly explained to different individuals. In particular, P09 emphasized that a significant portion of their time is spent reiterating basic principles of ML, such as how models function and why specific data requirements are necessary, to multiple stakeholders on a recurring basis. This continuous need for explanation effectively shifts their role toward ongoing instruction rather than project execution, consuming valuable time that could otherwise be dedicated to development work.
Although rework due to misalignment has been reported in prior work, in this setting, it is exacerbated by the need to repeatedly align stakeholders across both technical and domain-specific constraints, slowing iteration cycles.

The \textit{Unclear goals} (C6) challenge was noted by 4 individuals.
The expectations from both internal and external stakeholders regarding ML-enabled systems often lacked clarity, leading to misaligned priorities, resource allocation issues, and delays in project delivery. P04 elaborated that a common challenge involves addressing fundamental questions such as: \enquote{Why are we doing this?} and \enquote{Is this valuable for us?}. These are difficult but essential questions that require a clear declaration of the relevance and necessity of the work. Also, P04 explained that unclear goals are particularly evident in the need to justify foundational investments such as MLOps infrastructure. Substantial effort is required not only to build such platforms but also to continuously articulate their relevance to stakeholders. This includes explaining why these components are essential for project success, why progress may be constrained without them, and what risks may arise if they are not implemented. Hence, in the absence of clearly defined and shared objectives, teams must repeatedly defend strategic technical decisions rather than focus on execution. This ongoing need for justification reflects a broader misalignment in expectations, where the importance of enabling infrastructure is not fully understood.
While unclear goals are a common challenge in ML projects, in this context, they are closely tied to the need to justify investments in infrastructure, e.g., MLOps, within long-term, high-cost manufacturing processes.

A total of 4 participants mentioned \textit{\textbf{Employee shortage}} (C7) as a difficulty. There are not enough data scientists or ML engineers for some projects to divide and distribute tasks, and they therefore have to work on the complete pipeline. P09 stated that they were assigned a volume of project work that would realistically require multiple full-time contributors, highlighting a gap between organizational expectations and staffing levels. As a result, teams are often compelled to prioritize breadth over depth, focusing on developing numerous proof-of-concept solutions in order to address as many project demands as possible.
Although resource constraints are not totally unique to this domain, in our context, they are particularly impactful due to the need for highly specialized expertise and the limited availability of personnel capable of spanning both ML and domain-specific knowledge.

The \textit{Insufficient documentation} (C8) issue was raised by 4 participants. Inadequate and non-detailed documentation makes understanding and recreating ML projects harder. P01 described that documentation is often not sufficiently detailed to explain the rationale behind specific decisions. For example, why certain preprocessing steps were applied or why a particular ML architecture was chosen. Furthermore, P11 explained, working with a large, shared database that serves as the backend for multiple applications, containing extensive data but limited contextual explanation, is difficult. For instance, key variables, such as encoded measurement types, are represented by numerical values without clear documentation of their meaning, making it difficult to interpret or utilize the data correctly. As a result, identifying relevant data for a specific use case becomes a time-consuming and iterative process, requiring not only technical effort but also substantial domain expertise.
Documentation challenges are well known in prior studies, but here they are further complicated by the need to capture domain-specific knowledge and data semantics, which are often implicit and difficult to formalize.

Two participants reported encountering \textit{\textbf{Ignoring existing documentation}} (C9) as a challenge. People who want to integrate software sometimes do not read the ML requirements or specification documentation, which can lead to misalignment between the software and the ML models. P08 notes that documentation is primarily created as an internal resource, intended to support personal understanding or to assist close collaborators within the development team in navigating the system. In practice, however, it is perceived that external parties responsible for integrating the software rarely consult these materials.
This unique challenge is exacerbated by cross-team dependencies, where external stakeholders rely on documentation but may lack the context or incentives to consult it.

\textit{\textbf{Data governance complexity}} (C10) was acknowledged by 2 of the participants. The classification of data and data accessibility is not straightforward and makes communication less efficient. Additionally, rules concerning the sharing of data make accessing and looking for data more complicated. They also mentioned it is difficult to apply standard ML, like neural networks, because they have quite a small sample size, so they need a technique that is robust against small sample sizes. P01 stated that it is often unclear whether certain data can be stored in a cloud environment. In many cases, existing policies prohibit the use of cloud services altogether, which restricts access to a wide range of modern tools and platforms.
This challenge is more specific to the semiconductor domain, where strict regulations, data sensitivity, and infrastructure constraints significantly limit data accessibility and the applicability of standard ML practices.

Furthermore, 2 people pointed out \textit{\textbf{Unclear ongoing tasks}} (C11) as a problematic aspect. Team members do not know exactly what others are doing, so it might lead to confusion and inefficiency in the workflow. P10 mentioned it can be challenging to stay informed about ongoing activities in other departments.
Unlike general coordination issues reported in prior work, this unique challenge is closely tied to the matrix organization, where distributed teams and shifting project assignments reduce the transparency of ongoing activities.

\textit{\textbf{Employee tenure concerns}} (C12) was brought up by 1 interviewee during the study. Employees without a long-term position might decrease productivity because they are not properly integrated within the company, and it is unclear how long the project will continue. P12 mentioned that collaboration with a partner team located in a separate location, where financial instability and uncertain long-term prospects create additional strain. Although interpersonal interactions remain unaffected, the broader context introduces challenges in communication and coordination. Team members in this external group are perceived as not being fully integrated into the core department, which, combined with uncertainty about the project's continuity, contributes to a stressful working environment.
This issue appears less frequently in prior literature and is influenced by organizational and contractual structures, where temporary roles and external collaborations introduce uncertainty into long-term coordination.

According to 1 respondent, \textit{\textbf{Work overload}} (C13) was a concern. The workload for ML engineers is sometimes multiple times what one person can realistically handle, effectively requiring the effort of multiple individuals. This imbalance creates challenges, as engineers struggle to coordinate handovers, negotiate priorities, and distribute tasks across team members. P09 emphasizes that the imbalance between workload and available resources is a significant challenge in ML engineering practice. The technical deployment pipeline is described as highly automated, utilizing version control, pull requests, and continuous integration mechanisms to streamline model updates without requiring ongoing interaction with other teams. Specifically, the participant reports being assigned a volume of project work equivalent to several times their actual capacity, effectively requiring the effort of multiple engineers.
While workload imbalance is a general concern, in this context, it is intensified by the combination of high project demands and the need for end-to-end ownership in environments with limited specialized roles.

One person noted \textit{\textbf{ML trust issue}} (C14). Some people still do not believe in ML that can work better than humans in some aspects and areas. They think the results from ML models are worse than the results of human beings. This lack of trust creates challenges, as disagreements emerge over whether to rely on ML outputs, slowing decision-making and sometimes undermining team cohesion. P09 reports ongoing difficulties in convincing stakeholders of the reliability and effectiveness of their models, despite consistent performance in tasks such as segmentation. A particular point of contention arises from the model's tendency to identify more defect areas than human evaluators, which leads some stakeholders to conclude that the model performs worse.
Trust issues in this setting are heightened by the high precision and reliability requirements of manufacturing processes, where errors have significant consequences.

\textit{\textbf{Too many different communication channels}} (C15) was identified as a challenge by 1 participant. Multiple communication channels are open for discussing specific topics with different people. Over time, relevant information is shared across these channels, but later it becomes difficult to find because the exact channel where it was posted is forgotten. According to P04, working with collaboration platforms involves creating multiple channels to address specific topics and involve different stakeholders. While this structure facilitates targeted discussions, it also leads to the fragmentation of information over time. Important details shared in these channels can become difficult to retrieve, as individuals may no longer recall where specific information was originally posted. This lack of traceability creates inefficiencies and can hinder continuity in ongoing work.
The impact of these fragmented communication channels is amplified here due to the need for traceability and consistency across long-running, interdependent workflows.

The last challenge is \textit{\textbf{Missing technical leader}} (C16) mentioned by 1 participant. Without a technical leader, CoCo with external parties becomes difficult because no one has the combined technical oversight to evaluate trade-offs, translate requirements into feasible solutions, and ensure that design decisions remain consistent across teams. P12 emphasizes that, while individual contributors are often highly skilled, there is frequently no single person responsible for maintaining an overarching technical vision or ensuring coherence across components. In cross-organizational contexts, this lack of centralized technical oversight leads to recurring challenges, as no one effectively \enquote{holds all the strings together} to align decisions, manage dependencies, and guide the system's overall direction.
While leadership gaps can occur in the software-centric context, the absence of a central technical authority in the hardware-centric environment is particularly problematic given the need to align decisions across multiple disciplines and organizational boundaries.

To more clearly position our findings with respect to prior work, we distinguish between challenges that are well established in software-centric ML-enabled system development and how they manifest differently in the hardware-centric context. Several challenges we identified, such as unclear roles and duties, early-stage fragmented communication, ML knowledge deficiency, and rework, have been widely reported in prior studies~\cite{Nahar_2022, Mailach_2023, Busquim_2024}. However, our findings show that these challenges are not merely replicated but are intensified and reshaped by manufacturing constraints. For instance, role ambiguity in our context extends beyond typical interdisciplinary overlap and is reinforced by the matrix organizational structure, where responsibilities are distributed across departments with limited visibility. Similarly, early-stage communication issues are compounded by long development cycles and the need to align ML solutions with physical production processes, making misunderstandings more costly and harder to correct. ML knowledge gaps also manifest differently, as stakeholders must reason not only about software behavior but also about its interaction with precision-critical manufacturing steps.

In contrast, several challenges identified in our study appear more strongly rooted in the hardware-centric domain. These include data governance complexity, unclear ongoing tasks, and employee tenure concerns. For example, data governance complexity arises from strict regulations on production and metrology data, limiting accessibility and constraining the use of standard ML approaches. Likewise, unclear ongoing tasks and limited awareness of team members' skills are exacerbated by the matrix structure, where distributed expertise and shifting project assignments reduce transparency. Rather than claiming entirely new categories of challenges, our contribution lies in demonstrating how known CoCo issues are recontextualized under hardware, regulatory, and organizational constraints, and in identifying additional challenges that emerge from this specific setting.

\subsection{CoCo Approaches as Solutions}

\begin{figure*}[ht]
\centerline{\includegraphics[width=0.8\textwidth]{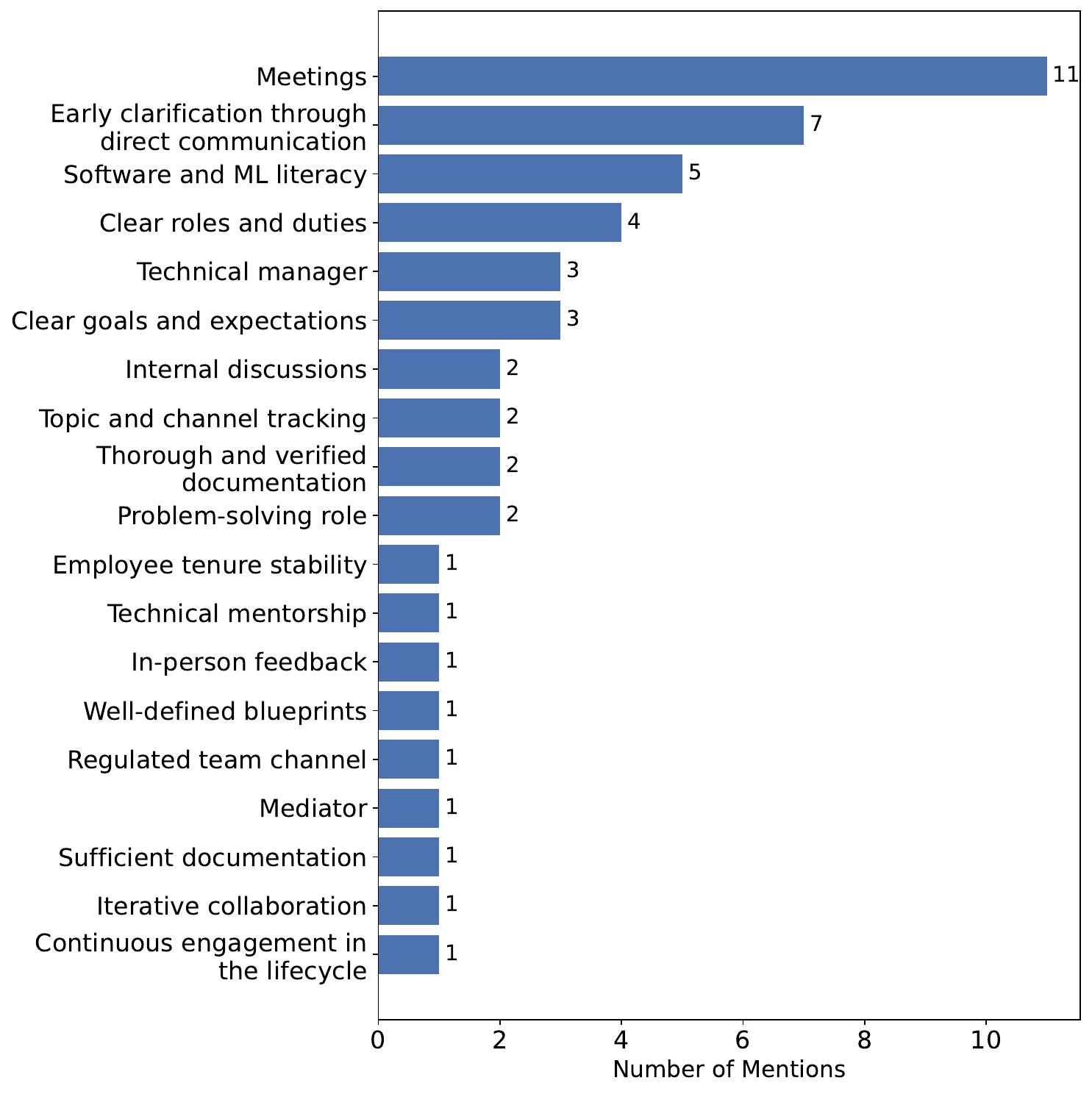}}
\caption{Solutions for CoCo challenges identified in the interviews with the number of mentions}
\label{fig3}
\end{figure*}

We asked participants directly about approaches they considered effective for overcoming CoCo challenges in building ML-enabled systems to answer \textbf{RQ3}. We identified 19 approaches, which are depicted in Fig~\ref{fig3}. The most frequently cited approach is \textit{Meetings - Scrum/Sprint meetings} to address C2, C4, C5, C6, C11. Regular meetings where participants with different roles talk about all processes of building ML-enabled systems and answer questions, e.g., about the data, how they process the data, which models they train, and how they do their workflow, can be highly beneficial for improving CoCo and understanding among team members. Additionally, showcasing the work biweekly by team members improves knowledge sharing and alignment across the team. Meetings on updates help to know about the changes and how the building of ML-enabled systems is going. P01 noted that the team divides the architecture work, with different members exploring distinct model architectures. They then hold a meeting to review updates and benchmark statistics, and collaboratively evaluate which model performs better or which architectural approach is more suitable. This approach fosters a deeper understanding of the models' strengths and weaknesses, encourages constructive feedback, and ensures that the team can collectively refine their strategies.

Seven participants identified \textit{Early clarification through direct communication} as a potential solution for C2, C5, and C6. Sufficient and effective CoCo among team members, especially in the initial phases, is crucial for ensuring that systems are developed efficiently and effectively. P07 outlined a project for which they initially underestimated the complexity of the task, assuming it required minimal effort and could be handled with a simple handover. They lacked a clear understanding of the requirements, and other departments were unaware of what they could actually deliver. However, a direct conversation proved to be extremely helpful as it allowed them to clarify expectations and bridge misunderstandings. What was originally estimated to take 100 hours of integration effort ultimately required only around 2 hours of actual work. The key difference was effective CoCo.

A total of 5 participants mentioned \textit{Software and ML literacy} as a possible remedy for C3. Educating technical team members about software engineering and ML capabilities enables them to make informed decisions about which ML models and techniques are best suited for specific tasks and identify potential challenges and limitations. For non-technical members, this knowledge helps them to understand the capabilities and limitations of ML-enabled systems. P06 noted that data scientists should possess at least a foundational understanding of software development practices, while software engineers and project managers should develop a basic grasp of statistics and ML concepts. Such cross-functional knowledge enables all stakeholders to better appreciate fundamental principles, such as the impact of data quality on model performance (\enquote{garbage in, garbage out}) and the limitations of data-driven methods when working with small or insufficient datasets.

\textit{Clear roles and duties} approach was suggested by 4 participants for C1, C5, C10. Clearly defined roles and responsibilities for team members prevent overlap and confusion, allowing them to focus on their specific tasks. This clarity helps improve efficiency, as team members can work without unnecessary interruptions or duplicated efforts. Additionally, it enhances accountability, as each person knows their expected contributions to the project. P02 indicated that while communication tools, e.g., Microsoft Teams, are generally sufficient for coordination, gaining a deep understanding of other team members' work, particularly in complex domains such as ML, requires significant time and effort. This highlights the need for well-defined responsibilities, as team members cannot realistically maintain in-depth knowledge of all aspects of a project. Clear role delineation, therefore, helps manage complexity, reduce unnecessary overhead, and ensure that expertise is effectively leveraged when needed.

Three respondents referred to \textit{Technical manager} as a viable approach for C10, C16. Having a technical manager or a manager who knows at least the basics about building ML-enabled systems, concerning getting access to data and merging different data sources, can better understand the challenges faced by the team and anticipate potential roadblocks. P06 emphasized that managers who have previously engaged, at least to some extent, in hands-on tasks such as integrating heterogeneous data sources or dealing with restricted or complex data access are better equipped to understand the practical difficulties involved in developing ML-enabled systems. This prior exposure enables them to contextualize technical constraints, make more informed decisions, and provide more effective support to their teams. Consequently, such experience was perceived as instrumental in bridging the gap between management and technical work, ultimately facilitating smoother project execution.

\textit{Clear goals and expectations} solution was proposed by 3 individuals for C6. A clear and shared goal that everyone can relate to. All parties should agree that the objectives are beneficial for everyone, ensuring alignment and commitment to delivering the intended outcome. P04 highlighted the value of dedicated goal-definition sessions, where all stakeholders align on the purpose and expected outcomes of a project. When team members collectively understand the underlying rationale (\enquote{why}) and anticipated benefits, they are more likely to develop a shared sense of direction and commitment. This mutual understanding was seen as facilitating smoother collaboration, as it provides a common foundation upon which different roles can coordinate their efforts and work toward agreed-upon objectives.

Two participants recommended \textit{Internal discussions} as a solution to the issue C2, C4, C6, and C11. There are internal discussions on project status reviews in talks, events, and knowledge sharing within the company to determine if it makes sense to proceed with a specific project, and whether they will reach their goal, especially for model architectures. P01 explained that, in their context, discussions primarily occur among data scientists. These discussions include assessing whether a given use case is suitable for ML, raising concerns about proposed solutions, and jointly reviewing progress and outcomes. Regular internal meetings were highlighted as particularly important for determining whether to continue or terminate a project based on its likelihood of success. Additionally, collaboration across teams often involves dividing work on different model architectures and later reconvening to compare performance metrics and benchmarks, enabling informed decisions about the most appropriate technical approach.

\textit{Topic and channel tracking} was a strategy noted by 2 participants for C15. As there are already multiple channels open for discussing specific topics with different people, a tracking system with notifications that can effectively track relevant shared information, details, communication, and the flow of information would be helpful and would help avoid wasted time. P04 suggested the need for supportive features, such as integrated tracking systems and notification mechanisms, that can help team members maintain visibility over relevant exchanges and better navigate distributed conversations. This approach was perceived as a way to significantly reduce time spent in unnecessary meetings, improve awareness of relevant discussions, and enhance overall communication efficiency within large organizations.

Two people brought up \textit{Thorough and verified documentation} during the discussions for C8. Writing good documentation with double-checks helps team members, especially people without prior knowledge of ML or data science, understand the system's functionality, data flow, and decision-making processes. P01 described an iterative documentation process in which different departments contribute in stages, with continuous feedback and validation cycles rather than isolated handovers. Initial guidelines and requirements are provided to structure the documentation, after which contributions are reviewed, extended, and refined collaboratively. This approach ensures that both technical and process-related aspects are adequately captured, including contextual explanations that may not be readily documented by all contributors.

\textit{Problem-solving role} was highlighted by 2 participants for C3. A consulting role for developers that checks the feasibility and meaningfulness of building an ML system for a specific use case, in the form of a community of practice, fosters collaboration and knowledge sharing among developers. P02 explained how communities of practice, such as those focused on ML or data science, provide accessible points of contact when teams encounter challenges beyond their immediate expertise. These communities facilitate ongoing exchange through regular meetings, where members present their work and share insights. As a result, such structures were perceived as fostering collaboration, enabling informed decision-making about ML use cases, and strengthening collective problem-solving capabilities.

\textit{Employee tenure stability} is a possible solution that emerged in the account of 1 participant for C12. Employees with long-term contracts and stability, especially external ones, tend to enhance productivity, as they are well-integrated within the company and provide continuity for ongoing projects. P12 contrasted the shared challenges of working with legacy systems, where difficulties are collectively experienced and do not create internal conflict, with the more complex dynamics arising from collaboration with external teams facing uncertain long-term prospects. This highlights that stable, long-term engagement, especially for external collaborators, supports better integration, reduces uncertainty, and fosters more effective and consistent collaboration over the course of a project.

A single participant pointed to \textit{Technical mentorship} as a relevant solution for C3, C16. Having a technical mentor ensures alignment, addresses challenges promptly, facilitates a unified vision, and connects all team members. While a technical manager primarily oversees resources and understands organizational or infrastructural constraints, a technical mentor focuses on guiding day-to-day technical direction and mentoring team members. While the P11 acknowledged that their manager possesses a solid understanding of ML and data-driven development, they emphasized that certain technical responsibilities, such as data modeling, storage, and platform infrastructure, are distributed across other specialized departments. In this context, a technical mentor was seen as particularly valuable for providing focused, day-to-day guidance within the team's domain, helping navigate dependencies on external platforms and ensuring alignment in technical decision-making. This role complements managerial functions by addressing practical implementation challenges and supporting team members in applying shared tools and infrastructures effectively within their specific area of expertise.

\textit{In-person feedback} was a strategy mentioned by 1 respondent for C2, C3, and C11. Observing a person evaluate the software in real time, seeing their expressions and how they interact with it, and receiving direct feedback in a live environment helps identify usability issues more effectively. P08 explained that observing users directly while they interact with the system, rather than through mediated channels, provides richer contextual insights, including non-verbal cues such as facial expressions and physical behavior. Such observations can reveal practical usability issues, for instance, when users physically move closer to the screen due to unclear interfaces or poor layout. This immediate and unfiltered feedback, obtained in the actual working environment, was perceived as more informative than remote alternatives, as it enables developers to better understand user experience and identify shortcomings that might otherwise remain unnoticed.

One participant proposed \textit{Well-defined blueprints} as a viable strategy for C5, C8, and C10. A well-defined blueprint for each use case within the company concerning data, ML models, and architectures that highlight the most promising models, architectures, and systems facilitates troubleshooting, scalability, and the reproducibility of results, and maintains high standards across teams. P01 mentioned that having well-documented blueprints or high-level descriptions of previously completed projects, including the models, architectures, and technologies used, as well as their outcomes, would significantly support teams when addressing similar use cases. Such guidelines would serve as a point of reference, enabling practitioners to identify promising approaches and make more informed design decisions. The availability of reusable templates and documented prior experiences was perceived as particularly valuable for establishing consistent ways of working, improving knowledge transfer, and reducing uncertainty when implementing new ML solutions.

One participant raised the \textit{Regulated team channel} approach for C2, C5, C6, C8, C11, and C15. A dedicated Microsoft Teams channel focused on active projects, providing updates on their current stage, documentation requirements, and responsible contributors, helps all team members align on project goals and have efficient communication. P01 recommended that, rather than relying on uncoordinated, individual exchanges between developers and external departments, communication be centralized through a dedicated channel managed with internal steering. In this setup, architects play a coordinating role by determining which projects are forwarded to other departments and ensuring that information is shared in a controlled and prioritized manner. The channel is used to provide updates on project status, current stages, documentation responsibilities, and assigned contributors. This structured approach was perceived as improving transparency and alignment, as all relevant participants are aware of how communication flows and can easily track project progress, in contrast to less organized, open communication formats that lacked clear visibility into project activities and stages.

\textit{Mediator/Middle communication layer} was considered a viable solution by 1 participant for C3, C6, and C10. A middle communication layer can bridge the CoCo gaps between employees at the company and external collaborators. P01 clarified that due to the involvement of multiple stakeholders and processes within a project, direct communication from individual developers to external collaborators would lead to fragmented exchanges and insufficient prioritization at the project level. Instead, communication is centralized through an internal structure involving architects and a dedicated channel, where decisions are made regarding which projects are communicated to external teams and how they progress. This mediated setup ensures that information about project stages, documentation responsibilities, and ongoing tasks is shared in a coordinated manner.

\textit{Sufficient documentation} was proposed by 1 individual for C8. The right amount of documentation is essential for ensuring transparency, reproducibility, and scalability. It also helps to recreate the projects when there is a change in data or a product. P01 explained that sufficient documentation is essential to ensure that projects can be reliably recreated. If there are changes in the data or the product, and they need to repeat a deployment or reproduce a specific component even a year later, the process must be reproducible. This requires that contributions from other departments are not only well documented but also understandable to team members who may not be ML experts, so that critical knowledge is preserved over time.

One participant identified \textit{Iterative collaboration} as a potential solution for C2, C5, and C11. Different groups work on their own tasks and send the final result to each other, and collaborate in multiple cycles to streamline the development process and ensure better outcomes. P01 noted that rather than following a linear, one-directional handover process, the work is carried out in multiple cycles, where one group contributes a portion, passes it on to another group for input, and then reviews and refines the result in subsequent iterations. This back-and-forth process enables continuous alignment between teams, allowing each party to incorporate feedback and add complementary knowledge from their respective domains.

\textit{Continuous engagement in the lifecycle} was suggested by a single participant for C2, C5, and C11. Active involvement of people who are engaged in building ML-enabled systems from the starting point to the end is crucial for ensuring the systems are designed effectively and function as intended. P01 mentioned that, in their context, a small group of data scientists remains responsible for a project without transitioning work to separate roles, such as ML engineers, on the same side. As a result, the same individuals oversee the process from the initial stages through to completion, ensuring continuity in knowledge, decisions, and responsibilities. This consistent involvement was perceived as beneficial for maintaining alignment across development phases, reducing information loss between handovers, and supporting a more coherent and efficient development process for ML-enabled systems.

\section{Discussion}

Unlike prior work that predominantly examines software-centric organizations where ML projects are largely data- and service-driven, our study deepens the understanding of CoCo challenges in MLE teams by focusing on a global semiconductor company using ML in its integrated manufacturing software. ML development in a global semiconductor company is tightly coupled with physical manufacturing processes and hardware constraints. ML projects must operate within strict reliability, precision, and traceability requirements, as errors can directly affect costly and irreversible production steps. Data availability is governed by sensor limitations, long production cycles, and stringent data governance policies, resulting in smaller, highly specialized datasets compared to the large and continuously generated datasets common in software companies. These constraints lead to longer development cycles, stronger dependencies between roles, and a greater need for cross-disciplinary collaboration than typically observed in software-centric ML projects. While prior studies have highlighted coordination challenges in ML-enabled systems~\citep{Amershi_2019, Nahar_2022}, they largely assume flexible data access and rapid iteration cycles, which contrasts with the constrained and tightly coupled environment observed in our study.

In the semiconductor industry, new products are introduced constantly and at a rapid pace. Therefore, priorities might change over time, and it is always necessary to adapt. Additionally, according to participants of the study, this global manufacturing company is not like a car manufacturer that produces millions of cars each year. There are relatively small production numbers but a high-tech process that can cause numerous problems because of the very rigid and complex requirements that are on the edge of today's physics, which is why the work is complex, and the CoCo is so significant.

One of the central insights from this study is that CoCo should be treated as a first-class engineering concern rather than just a soft issue. When engineers spend weeks resolving misaligned expectations, when data scientists carry the burden of entire end-to-end pipelines, or when production delays stem from insufficient documentation, the consequences of suboptimal CoCo practices are just as impactful as poor algorithmic choices. This perspective complements recent calls in ML-enabled systems research to account for socio-technical dimensions~\citep{Mailach_2023, Busquim_2024}, but extends them into the underexplored domain of semiconductor manufacturing, where the stakes of CoCo breakdowns are especially high due to the cost and irreversibility of production processes. In prior work~\citep{Nahar_2022, Busquim_2024, Piorkowski_2021}, CoCo challenges are often discussed in terms of their impact on development efficiency, coordination overhead, or delays in delivery. In contrast, our findings show that in hardware-centric environments, these issues can directly affect production reliability and introduce operational risks.

Participants described that coordination is often constrained by factors such as fragmented communication across departments, uneven ML literacy, and unclear role boundaries. These challenges extend beyond tooling or workflow design and reflect deeper organizational characteristics, including the matrix structure, distributed expertise, and rigid process dependencies typical of large manufacturing environments. Understanding these mechanisms is essential for designing collaboration strategies that integrate human, organizational, and technical dimensions more effectively. This fragmentation creates persistent communication gaps, similar to those discussed in \cite{Busquim_2024, Almahmoud_2021}, where differing priorities and vocabularies lead to friction. However, unlike prior studies where such gaps primarily affect development efficiency, in our context, they are further constrained by interdependencies between software and physical processes, making misalignment more difficult to resolve. Our interviews also underscore that these gaps are not easily closed by tools alone; rather, they demand ongoing negotiation of shared goals and responsibilities, especially at key project stages such as data acquisition and model deployment.

A promising research and practice direction may be to embed CoCo metrics into project evaluation, such as tracking communication load, meeting effectiveness, or cross-team dependency resolution. In the same way that teams routinely track code quality and model accuracy, they should also assess the quality of collaboration, making it a visible and accountable part of socio-technical MLE work.

Furthermore, the role ambiguity we observed, where mostly data scientists act as ML engineers, pipeline developers, and communicators of ML concepts, reflects a structural imbalance. On the one hand, such generalist behavior creates agility in a matrixed organization, but on the other hand, it blurs accountability and produces inefficiencies. Our findings suggest that the issue is not role overlap itself, but the absence of clearly defined responsibilities per project. This finding aligns with recent work by~\cite{Nahar_2022, Wu_2024, Nazir_2024}, which identified role ambiguity and responsibility confusion as recurring problems in ML-enabled systems. However, in our context, this issue is often compounded by the limited availability of specialized personnel.
Rather than enforcing strict role boundaries, deliberate coordination mechanisms and boundary objects, such as shared documentation templates and responsibility matrices, can help teams maintain clarity while preserving flexibility.

The new CoCo challenges, such as unclear ongoing tasks, limited awareness of team members' skills, and ML trust issues, are identified through this interview study in such industries that can hinder productivity, reduce collaboration efficiency, and lead to miscommunication. These difficulties are closely tied to the company's matrix organization. While this structure enables flexibility by drawing in expertise dynamically, it also diffuses accountability and creates ambiguity in ownership. In line with prior work by~\cite{Nahar_2022, Suresh_2021} on coordination in complex organizations, we argue that these problems are not incidental but structurally rooted in the organizational form itself. Compared to previous studies, these challenges are less about tool misuse and more about visibility and coordination across organizational boundaries.

Another contribution of our study is the identification of a variety of CoCo approaches that practitioners adopt in response to specific challenges, such as unclear goals, misaligned expectations, or role ambiguity across teams. For instance, sprint planning meetings and structured in-person feedback sessions were described as ways to resolve misunderstandings about scope and deliverables. Similarly, initiatives to improve software and ML literacy were seen as addressing knowledge gaps between domain experts and technical staff, while communities of practice provided a forum for resolving isolated technical problems and sharing experiences across projects. These practices represent important bottom-up strategies that help practitioners navigate structural frictions in the absence of a centralized MLE team. While similar practices have been reported in prior work by~\cite{Busquim_2024}, our findings suggest they play a more critical compensatory role in environments lacking centralized coordination structures. Our analysis also suggests that while these measures are useful, they are often reactive and situational rather than systemic. To more effectively mitigate recurring CoCo issues in a matrix environment, deliberate mechanisms such as lightweight responsibility definitions, intermediary coordination roles, or rotating leads could be introduced.

While many of the identified CoCo challenges, such as unclear roles, early-stage fragmented communication, and ML knowledge gaps, are consistent with findings from prior research in software-centric organizations~\citep{Nahar_2022, Busquim_2024}, several issues appear contextually shaped by the semiconductor domain. In particular, challenges related to data governance complexity and work overload are tightly linked to the constraints of working with highly sensitive production and metrology data and long hardware development cycles. These factors restrict data access, complicate experimentation, and make CoCo across departments considerably more difficult than in purely digital product settings. This contrasts with software-centric environments, where data accessibility and rapid experimentation are often assumed as given. Similarly, limited awareness of team members' skills and unclear ongoing tasks stem partly from the company's matrix structure. On the solutions side, practices like mediator or middle communication layers, communities of practice, and well-defined blueprints for ML use cases reflect adaptive mechanisms tailored to this industrial context. Nevertheless, approaches such as early clarification through direct communication and clear role and responsibility definitions are likely transferable to other domains, supporting prior findings that some CoCo practices generalize across contexts.

Although our study identified a range of CoCo challenges and approaches, not all issues currently have clear or actionable solutions proposed by practitioners. Several challenges, e.g., ignoring existing documentation and ML trust issues, were recognized as obstacles without systematic mitigation strategies in place. In the same vein, challenges like work overload were acknowledged but lacked concrete organizational responses, suggesting that these areas remain underexplored in current practice. This highlights that while some CoCo challenges already have some suggestions to be addressed through practical measures and process improvements, others remain insufficiently tackled and call for further research and organizational experimentation to develop effective and sustainable solutions.

In addition, we discussed the various tools and styles practitioners use for documentation in this company. However, there are still issues related to different types of documentation. For example, while the available documentation typically provides a high-level overview of the system, it rarely captures the underlying reasoning behind each design choice. This lack of detailed information can make it difficult for (new) team members to understand the rationale behind certain decisions~\citep{Mitchell_2019, Gebru_2021, Retzlaff_2024}. This finding is consistent with prior work on documentation debt in ML systems, but is further complicated in our context by domain-specific data semantics and cross-team dependencies. Hence, attention to verified and thorough documentation is essential at each phase of building ML-enabled systems~\citep{Pineau_2021, Bhat_2023}, such as a decision tracker.

A direct comparison between software-centric and hardware-centric ML environments reveals both strong similarities and important differences in CoCo dynamics. On the one hand, many challenges identified in prior software-focused studies, such as role ambiguity, interdisciplinary communication gaps, and insufficient documentation, are also present in our setting, confirming that these are fundamental issues in ML-enabled systems development. On the other hand, their manifestation differs significantly under manufacturing constraints. In software-centric environments, these challenges primarily affect development efficiency, iteration speed, and maintainability, as teams typically operate with flexible data access, rapid feedback loops, and fewer physical dependencies. In contrast, in the semiconductor context, the same challenges are amplified by strict data governance policies, long production cycles, and tight coupling between software and physical processes. As a result, miscommunication or unclear responsibilities slow down development and can directly impact production reliability, delay costly manufacturing steps, and increase operational risk. Furthermore, the highly specialized expertise distribution reduces visibility across teams, making coordination more dependent on organizational mechanisms rather than tooling alone. These differences suggest that while core CoCo challenges are shared across domains, their consequences, severity, and mitigation strategies are strongly shaped by the underlying industrial context.

Overall, our study shows that CoCo in MLE teams is about improving day-to-day practices and rethinking how coordination principles are adapted to fit the realities of matrix organizations. This suggests that practitioners can continue to refine bottom-up approaches to mitigate ambiguity, while organizations should experiment with structural interventions that bind clarity and responsibility more firmly across projects. By situating employee experiences within this broader perspective, we contribute both a grounded understanding of CoCo challenges and a conceptual basis for imagining how they might be more effectively addressed.

This study also has practical implications for both practitioners and researchers. For practitioners, the identified approaches offer actionable starting points for improving MLE team CoCo in settings where traditional software engineering practices may not directly apply. For researchers, our findings raise new questions about how socio-technical practices evolve in resource-constrained, interdisciplinary environments and how organizational context shapes the adoption of CoCo strategies. More broadly, our work complements existing literature by showing how established challenges persist but evolve under hardware-centric constraints.

\section{Threats to Validity}

To contextualize our findings, we discuss potential threats to the validity of this study in different dimensions. While the research provides valuable insights into CoCo challenges within a complex industrial setting, several factors may affect the interpretation, generalizability, and consistency of the results.

\subsection{Construct Validity}

We pilot-tested our interview protocol and designed the semi-structured interview questions based on prior literature in software engineering and ML about CoCo-related topics to ensure alignment with established definitions. However, concepts like \enquote{misunderstandings} or \enquote{communication breakdown} may be interpreted differently by participants.
Moreover, to minimize bias during analysis, we employed investigator triangulation. The first two authors independently coded the transcripts via open coding, followed by iterative discussions to resolve disagreements and develop the final coding scheme. Because interview transcripts and extracted passages cannot be shared due to organizational confidentiality, readers cannot directly inspect the analytical traces. We mitigate this limitation by providing detailed descriptions of the coding process and representative indirect anonymized interviewees' quotes.

\subsection{Internal Validity}

Participants might alter their responses based on perceived study goals, but we avoided disclosing specific research hypotheses to participants and framed the study broadly to cope with this threat. Changes in interview protocols, question phrasing, or interviewer behavior across sessions could lead to inconsistent data collection, but we used a standardized interview guide and pilot-tested interviews to maintain consistency.
Moreover, while qualitative interview studies can in principle be influenced by internal dynamics, role hierarchies, or the presence of a company representative at the beginning of interviews, we have no indication that this was a significant factor in our context. Participants were explicitly encouraged to speak openly, and their accounts consistently included discussions of both strengths and weaknesses in current practices.

Furthermore, employees may conceal or ignore negative opinions about internal CoCo due to fear of professional repercussions or loyalty to the company. This can lead to biased responses that do not accurately reflect underlying issues. However, we estimate that its effect on the validity of our findings is limited, as participants often provided candid examples and constructive criticism that reflected both strengths and weaknesses in current practices.

\subsection{External Validity}

Our participants may not be fully representative of broader MLE teams. Differences in seniority or experience may result in divergent perspectives on CoCo challenges.
Besides, the subject company is a highly specialized entity with unique domain characteristics, with interdisciplinary collaboration among physicists, data, process, and software engineering. Despite the company's presence in more than 50 countries, our findings may not apply to companies with different domain constraints. We clearly define the study's scope and provide rich contextual details about the company's environment to help readers assess transferability and compare findings to related studies in similar domains.
As a case study of one company, it may be difficult to determine which observed challenges are specific to this firm's organizational culture versus those common across the industry.

\subsection{Reliability}

Inconsistent coding of qualitative data or a lack of agreement among authors can undermine the reliability of conclusions. To mitigate this, we implemented a rigorous coding framework and conducted regular calibration sessions to ensure consistency among the authors.
A small or non-representative subset of employees, such as only senior software engineers or members of a single team, can produce results that fail to reflect broader organizational challenges. To alleviate this problem, we made sure that participants represented a diverse range of roles and departments.

\section{Conclusion}

As the application of ML spreads into manufacturing environments such as the semiconductor industry, it becomes increasingly important to understand and address the CoCo challenges faced by MLE teams. This study contributes a grounded perspective on CoCo in ML-enabled system development, informed by the lived experiences of practitioners at a global semiconductor company using ML in its integrated manufacturing software. By identifying 16 core challenges, most notably unclear roles and responsibilities, and highlighting 19 collaborative approaches used in practice, we provide a nuanced account of the organizational challenges and practices that shape MLE work in hardware-centric environments.

Our findings emphasize that the success of ML initiatives depends not only on technical excellence but also on the clarity of team structures and the effectiveness of cross-functional CoCo in matrixed organizations. These insights offer both practical recommendations for teams navigating similar contexts and a foundation for future research aimed at designing tools, processes, and organizational models that better support interdisciplinary ML work. These insights provide a contextualized account of ML CoCo in a semiconductor setting and offer a starting point for future comparative studies across industries.

\section*{CRediT authorship contribution statement}

\textbf{Aidin Azamnouri:} Writing – original draft, Writing – review \& editing, Visualization, Validation, Methodology, Investigation, Formal analysis, Data curation, Conceptualization. \textbf{Markus Haug:} Writing – review \& editing, Validation, Methodology, Investigation, Formal analysis, Data curation, Conceptualization. \textbf{Lucas Woltmann:} Writing – review \& editing, Validation, Conceptualization. \textbf{Manuel Fritz:} Writing – review \& editing, Validation. \textbf{Justus Bogner:} Writing – review \& editing, Validation. \textbf{Stefan Wagner:} Writing – review \& editing, Validation.

\section*{Statement on open data and ethics}

All artifacts containing the interview preamble, guide, and aggregated analysis can be found at \url{https://doi.org/10.5281/zenodo.17358331}.
Due to the organization's privacy considerations and data protection regulations, the interview transcripts and extractions cannot be made publicly available. It is not permitted for us to distribute or disclose this information.

Regarding ethics, a comprehensive privacy statement was developed to ensure that all participants were clearly informed about how their personal data would be collected, processed, stored, and protected throughout the course of the study. This statement outlined the measures taken to safeguard confidentiality and to comply with applicable data protection regulations. In parallel, a detailed informed consent form was prepared to provide participants with a clear understanding of the study's purpose, procedures, and their rights, including the voluntary nature of participation and the option to withdraw at any time without consequence.

Prior to initiating participant recruitment, the complete study protocol, along with all supporting materials, including the privacy statement and consent documentation, was reviewed by the company's employee representative committee (workers' council). This review ensured that the study adhered to relevant ethical guidelines, legal requirements, and organizational policies governing employee participation in research activities.

Following this approval, all prospective participants were provided with a comprehensive study information package prior to their interviews. This package included an overview of the study, detailed privacy information, and the informed consent form. Participants were asked to carefully review these materials and to provide their signed consent prior to participation. This process was designed to ensure that all participants were fully informed and had sufficient time to consider their involvement before engaging in the study.

\section*{Declaration of competing interest}
The authors declare that they have no known competing financial interests or personal relationships that could have appeared to influence the work reported in this paper.

\section*{Acknowledgements}

The project on which this work is based was sponsored by the Federal Ministry of Education and Research (BMBF) under the funding code 01IS23069. We would like to express our sincere gratitude to all those who contributed to the success of this work, especially Christian Schriever (Zeiss) and Marvin Muñoz Barón.

\bibliographystyle{cas-model2-names}
\bibliography{cas-refs}

@inproceedings{Sculley_2015,
 author = {Sculley, D. and Holt, Gary and Golovin, Daniel and Davydov, Eugene and Phillips, Todd and Ebner, Dietmar and Chaudhary, Vinay and Young, Michael and Crespo, Jean-Fran\c{c}ois and Dennison, Dan},
 booktitle = {Advances in Neural Information Processing Systems},
 editor = {C. Cortes and N. Lawrence and D. Lee and M. Sugiyama and R. Garnett},
 pages = {},
 publisher = {Curran Associates, Inc.},
 title = {Hidden Technical Debt in Machine Learning Systems},
 url = {https://proceedings.neurips.cc/paper_files/paper/2015/file/86df7dcfd896fcaf2674f757a2463eba-Paper.pdf},
 volume = {28},
 year = {2015}
}

@inproceedings{Amershi_2019,
title={Software Engineering for Machine Learning: A Case Study},
url={http://dx.doi.org/10.1109/icse-seip.2019.00042},
DOI={10.1109/icse-seip.2019.00042},
booktitle={2019 IEEE/ACM 41st International Conference on Software Engineering: Software Engineering in Practice (ICSE-SEIP)},
publisher={IEEE},
author={Amershi, Saleema and Begel, Andrew and Bird, Christian and DeLine, Robert and Gall, Harald and Kamar, Ece and Nagappan, Nachiappan and Nushi, Besmira and Zimmermann, Thomas},
year={2019},
month=may,
pages={291–300}
}

@article{Wan_2020,
title={How does Machine Learning Change Software Development Practices?},
ISSN={2326-3881},
url={http://dx.doi.org/10.1109/tse.2019.2937083},
DOI={10.1109/tse.2019.2937083},
journal={IEEE Transactions on Software Engineering},
publisher={Institute of Electrical and Electronics Engineers (IEEE)},
author={Wan, Zhiyuan and Xia, Xin and Lo, David and Murphy, Gail C.},
year={2020},
pages={1–1}
}

@inproceedings{Honkanen_2022,
series={AHFE 2022},
title={Multidisciplinary Teamwork in Machine Learning Operations (MLOps)}, ISSN={2771-0718},
url={http://dx.doi.org/10.54941/ahfe1002261},
DOI={10.54941/ahfe1002261},
booktitle={Human Factors, Business Management and Society},
publisher={AHFE International},
author={Honkanen, Tapani and Odwyer, Jonny and Salminen, Vesa},
year={2022},
collection={AHFE 2022}
}

@inproceedings{Sambasivan_2021,
series={CHI ’21},
title={“Everyone wants to do the model work, not the data work”: Data Cascades in High-Stakes AI},
url={http://dx.doi.org/10.1145/3411764.3445518},
DOI={10.1145/3411764.3445518},
booktitle={Proceedings of the 2021 CHI Conference on Human Factors in Computing Systems},
publisher={ACM},
author={Sambasivan, Nithya and Kapania, Shivani and Highfill, Hannah and Akrong, Diana and Paritosh, Praveen and Aroyo, Lora M},
year={2021},
month=may,
pages={1–15},
collection={CHI ’21}
}

@inproceedings{Mailach_2023,
title={Socio-Technical Anti-Patterns in Building ML-Enabled Software: Insights from Leaders on the Forefront},
url={http://dx.doi.org/10.1109/icse48619.2023.00067},
DOI={10.1109/icse48619.2023.00067},
booktitle={2023 IEEE/ACM 45th International Conference on Software Engineering (ICSE)},
publisher={IEEE},
author={Mailach, Alina and Siegmund, Norbert},
year={2023},
month=may,
pages={690–702}
}

@article{Annunziata_2025,
title={Uncovering Community Smells in Machine Learning-Enabled Systems: Causes, Effects, and Mitigation Strategies},
volume={34},
ISSN={1557-7392},
url={http://dx.doi.org/10.1145/3712198},
DOI={10.1145/3712198},
number={6},
journal={ACM Transactions on Software Engineering and Methodology}, publisher={Association for Computing Machinery (ACM)},
author={Annunziata, Giusy and Lambiase, Stefano and Tamburri, Damian A. and van den Heuvel, Willem-Jan and Palomba, Fabio and Catolino, Gemma and Ferrucci, Filomena and De Lucia, Andrea},
year={2025},
month=jul,
pages={1–48}
}

@article{Zaharia_2018,
  title={Accelerating the machine learning lifecycle with MLflow.},
  author={Zaharia, Matei and Chen, Andrew and Davidson, Aaron and Ghodsi, Ali and Hong, Sue Ann and Konwinski, Andy and Murching, Siddharth and Nykodym, Tomas and Ogilvie, Paul and Parkhe, Mani and others},
  journal={IEEE Data Eng. Bull.},
  volume={41},
  number={4},
  pages={39--45},
  year={2018}
}

@inproceedings{Hove_2005,
title={Experiences from Conducting Semi-structured Interviews in Empirical Software Engineering Research},
url={http://dx.doi.org/10.1109/metrics.2005.24},
DOI={10.1109/metrics.2005.24},
booktitle={11th IEEE International Software Metrics Symposium (METRICS’05)},
publisher={IEEE},
author={Hove, S.E. and Anda, B.},
year={2005},
pages={23–23}
}

@article{Runeson_2008,
title={Guidelines for conducting and reporting case study research in software engineering},
volume={14},
ISSN={1573-7616},
url={http://dx.doi.org/10.1007/s10664-008-9102-8},
DOI={10.1007/s10664-008-9102-8},
number={2},
journal={Empirical Software Engineering},
publisher={Springer Science and Business Media LLC},
author={Runeson, Per and Höst, Martin},
year={2008},
month=dec,
pages={131–164}
}

@inproceedings{Nahar_2022,
series={ICSE ’22},
title={Collaboration challenges in building ML-enabled systems: communication, documentation, engineering, and process},
url={http://dx.doi.org/10.1145/3510003.3510209},
DOI={10.1145/3510003.3510209},
booktitle={Proceedings of the 44th International Conference on Software Engineering},
publisher={ACM},
author={Nahar, Nadia and Zhou, Shurui and Lewis, Grace and Kästner, Christian},
year={2022},
month=may,
pages={413–425},
collection={ICSE ’22}
}

@article{Haberl_2024,
title={Take the aTrain. Introducing an interface for the Accessible Transcription of Interviews},
volume={41},
ISSN={2214-6350},
url={http://dx.doi.org/10.1016/j.jbef.2024.100891},
DOI={10.1016/j.jbef.2024.100891},
journal={Journal of Behavioral and Experimental Finance},
publisher={Elsevier BV},
author={Haberl, Armin and Fleiß, Jürgen and Kowald, Dominik and Thalmann, Stefan},
year={2024},
month=mar,
pages={100891}
}

@article{Seaman_1999,
title={Qualitative methods in empirical studies of software engineering}, volume={25},
ISSN={2326-3881},
url={http://dx.doi.org/10.1109/32.799955},
DOI={10.1109/32.799955},
number={4},
journal={IEEE Transactions on Software Engineering},
publisher={Institute of Electrical and Electronics Engineers (IEEE)},
author={Seaman, C.B.},
year={1999},
month=jul,
pages={557–572}
}

@incollection{Adams_2015,
title={Conducting Semi‐Structured Interviews},
booktitle={Handbook of Practical Program Evaluation},
ISBN={9781119171386},
url={http://dx.doi.org/10.1002/9781119171386.ch19},
DOI={10.1002/9781119171386.ch19},
journal={Handbook of Practical Program Evaluation},
publisher={Wiley},
author={Adams, William C.},
year={2015},
month=aug,
pages={492–505}
}

@inproceedings{Cruzes_2011,
title={Recommended Steps for Thematic Synthesis in Software Engineering},
url={http://dx.doi.org/10.1109/esem.2011.36},
DOI={10.1109/esem.2011.36},
booktitle={2011 International Symposium on Empirical Software Engineering and Measurement},
publisher={IEEE},
author={Cruzes, D. S. and Dyba, T.},
year={2011}, 
month=sep,
pages={275–284}
}

@inproceedings{Azamnouri_2025,
title={CoCo Challenges in ML Engineering Teams: How to Collaboratively Build ML-Enabled Systems},
url={http://dx.doi.org/10.1109/cain66642.2025.00036},
DOI={10.1109/cain66642.2025.00036},
booktitle={2025 IEEE/ACM 4th International Conference on AI Engineering – Software Engineering for AI (CAIN)},
publisher={IEEE},
author={Azamnouri, Aidin},
year={2025},
month=apr,
pages={241–243}
}

@inbook{Busquim_2024,
title={On the Interaction Between Software Engineers and Data Scientists When Building Machine Learning-Enabled Systems},
ISBN={9783031562815},
ISSN={1865-1356},
url={http://dx.doi.org/10.1007/978-3-031-56281-5_4},
DOI={10.1007/978-3-031-56281-5_4},
booktitle={Software Quality as a Foundation for Security}, publisher={Springer Nature Switzerland},
author={Busquim, Gabriel and Villamizar, Hugo and Lima, Maria Julia and Kalinowski, Marcos},
year={2024},
pages={55–75}
}

@inproceedings{Busquim_2024_2,
series={SBES 2024},
title={Towards Effective Collaboration between Software Engineers and Data Scientists developing Machine Learning-Enabled Systems},
url={http://dx.doi.org/10.5753/sbes.2024.3027},
DOI={10.5753/sbes.2024.3027},
booktitle={Anais do XXXVIII Simpósio Brasileiro de Engenharia de Software (SBES 2024)},
publisher={Sociedade Brasileira de Computação},
author={Busquim, Gabriel and Araújo, Allysson Allex and Lima, Maria Julia and Kalinowski, Marcos},
year={2024},
month=sep,
pages={24–34},
collection={SBES 2024}
}

@article{Piorkowski_2021,
title={How AI Developers Overcome Communication Challenges in a Multidisciplinary Team: A Case Study},
volume={5},
ISSN={2573-0142},
url={http://dx.doi.org/10.1145/3449205},
DOI={10.1145/3449205},
number={CSCW1},
journal={Proceedings of the ACM on Human-Computer Interaction},
publisher={Association for Computing Machinery (ACM)},
author={Piorkowski, David and Park, Soya and Wang, April Yi and Wang, Dakuo and Muller, Michael and Portnoy, Felix},
year={2021},
month=apr,
pages={1–25}
}

@article{Almahmoud_2021,
title={How Teams Communicate about the Quality of ML Models: A Case Study at an International Technology Company},
volume={5},
ISSN={2573-0142},
url={http://dx.doi.org/10.1145/3463934},
DOI={10.1145/3463934},
number={GROUP},
journal={Proceedings of the ACM on Human-Computer Interaction},
publisher={Association for Computing Machinery (ACM)},
author={Almahmoud, Jumana and DeLine, Robert and Drucker, Steven M.},
year={2021},
month=jul,
pages={1–24}
}

@inproceedings{Wu_2024,
series={CAIN 2024},
title={An Exploratory Study of V-Model in Building ML-Enabled Software: A Systems Engineering Perspective},
url={http://dx.doi.org/10.1145/3644815.3644951},
DOI={10.1145/3644815.3644951},
booktitle={Proceedings of the IEEE/ACM 3rd International Conference on AI Engineering - Software Engineering for AI},
publisher={ACM},
author={Wu, Jie JW},
year={2024},
month=apr,
pages={30–40},
collection={CAIN 2024}
}

@article{Nazir_2024,
title={Architecting ML-enabled systems: Challenges, best practices, and design decisions},
volume={207},
ISSN={0164-1212},
url={http://dx.doi.org/10.1016/j.jss.2023.111860},
DOI={10.1016/j.jss.2023.111860},
journal={Journal of Systems and Software},
publisher={Elsevier BV},
author={Nazir, Roger and Bucaioni, Alessio and Pelliccione, Patrizio},
year={2024},
month=jan,
pages={111860}
}

@article{Krause_J_ttler_2022,
title={Interdisciplinary Collaborations in Digital Health Research: Mixed Methods Case Study},
volume={9},
ISSN={2292-9495},
url={http://dx.doi.org/10.2196/36579},
DOI={10.2196/36579},
number={2},
journal={JMIR Human Factors},
publisher={JMIR Publications Inc.},
author={Krause-Jüttler, Grit and Weitz, Jürgen and Bork, Ulrich},
year={2022},
month=may,
pages={e36579}
}

@inproceedings{Lima_2022,
title={MLOps: Practices, Maturity Models, Roles, Tools, and Challenges – A Systematic Literature Review},
url={http://dx.doi.org/10.5220/0010997300003179},
DOI={10.5220/0010997300003179},
booktitle={Proceedings of the 24th International Conference on Enterprise Information Systems},
publisher={SCITEPRESS - Science and Technology Publications},
author={Lima, Anderson and Monteiro, Luciano and Furtado, Ana},
year={2022}
}

@article{Polyzotis_2018,
title={Data Lifecycle Challenges in Production Machine Learning: A Survey},
volume={47},
ISSN={0163-5808},
url={http://dx.doi.org/10.1145/3299887.3299891},
DOI={10.1145/3299887.3299891},
number={2},
journal={ACM SIGMOD Record},
publisher={Association for Computing Machinery (ACM)},
author={Polyzotis, Neoklis and Roy, Sudip and Whang, Steven Euijong and Zinkevich, Martin},
year={2018},
month=dec,
pages={17–28}
}

@article{Recupito_2024,
title={Technical debt in AI-enabled systems: On the prevalence, severity, impact, and management strategies for code and architecture},
volume={216},
ISSN={0164-1212},
url={http://dx.doi.org/10.1016/j.jss.2024.112151},
DOI={10.1016/j.jss.2024.112151},
journal={Journal of Systems and Software},
publisher={Elsevier BV},
author={Recupito, Gilberto and Pecorelli, Fabiano and Catolino, Gemma and Lenarduzzi, Valentina and Taibi, Davide and Di Nucci, Dario and Palomba, Fabio},
year={2024},
month=oct,
pages={112151}
}

@inproceedings{Lewis_2021,
title={Characterizing and Detecting Mismatch in Machine-Learning-Enabled Systems},
url={http://dx.doi.org/10.1109/wain52551.2021.00028},
DOI={10.1109/wain52551.2021.00028},
booktitle={2021 IEEE/ACM 1st Workshop on AI Engineering - Software Engineering for AI (WAIN)},
publisher={IEEE},
author={Lewis, Grace A. and Bellomo, Stephany and Ozkaya, Ipek},
year={2021},
month=may,
pages={133–140}
}

@article{Eken_2025,
title={A Multivocal Review of MLOps Practices, Challenges and Open Issues},
volume={58},
ISSN={1557-7341},
url={http://dx.doi.org/10.1145/3747346},
DOI={10.1145/3747346},
number={2},
journal={ACM Computing Surveys},
publisher={Association for Computing Machinery (ACM)},
author={Eken, Beyza and Pallewatta, Samodha and Tran, Nguyen and Tosun, Ayse and Babar, Muhammad Ali},
year={2025},
month=sep,
pages={1–35}
}

@article{Xu_2024,
title={A fast ramp-up framework for wafer yield improvement in semiconductor manufacturing systems},
volume={76},
ISSN={0278-6125},
url={http://dx.doi.org/10.1016/j.jmsy.2024.07.001},
DOI={10.1016/j.jmsy.2024.07.001},
journal={Journal of Manufacturing Systems},
publisher={Elsevier BV},
author={Xu, Hong-Wei and Zhang, Qi-Hua and Sun, Yan-Ning and Chen, Qun-Long and Qin, Wei and Lv, You-Long and Zhang, Jie},
year={2024},
month=oct,
pages={222–233}
}

@article{Li_2023,
title={Reinforcement learning for process control with application in semiconductor manufacturing},
volume={56},
ISSN={2472-5862},
url={http://dx.doi.org/10.1080/24725854.2023.2219290},
DOI={10.1080/24725854.2023.2219290},
number={6},
journal={IISE Transactions},
publisher={Informa UK Limited},
author={Li, Yanrong and Du, Juan and Jiang, Wei},
year={2023},
month=jul,
pages={585–599}
}

@inproceedings{Suresh_2021,
series={CHI ’21},
title={Beyond Expertise and Roles: A Framework to Characterize the Stakeholders of Interpretable Machine Learning and their Needs},
url={http://dx.doi.org/10.1145/3411764.3445088},
DOI={10.1145/3411764.3445088},
booktitle={Proceedings of the 2021 CHI Conference on Human Factors in Computing Systems},
publisher={ACM},
author={Suresh, Harini and Gomez, Steven R. and Nam, Kevin K. and Satyanarayan, Arvind},
year={2021},
month=may,
pages={1–16},
collection={CHI ’21}
}

@article{Bhatt_2020,
author={Bhatt, Umang and Andrus, McKane and Weller, Adrian and Xiang, Alice},
title={Machine Learning Explainability for External Stakeholders},
journal={CoRR},
volume={abs/2007.05408},
year={2020},
url={https://arxiv.org/abs/2007.05408},
DOI={10.48550/arXiv.2007.05408},
}

@article{Pineau_2021,
author={Joelle Pineau and Philippe Vincent-Lamarre and Koustuv Sinha and Vincent Lariviere and Alina Beygelzimer and Florence d'Alche-Buc and Emily Fox and Hugo Larochelle},
title={Improving Reproducibility in Machine Learning Research(A Report from the NeurIPS 2019 Reproducibility Program)},
journal={Journal of Machine Learning Research},
year={2021},
volume={22},
number={164},
pages={1--20},
url={http://jmlr.org/papers/v22/20-303.html}
}

@inproceedings{Bhat_2023,
series={CHI ’23},
title={Aspirations and Practice of ML Model Documentation: Moving the Needle with Nudging and Traceability},
url={http://dx.doi.org/10.1145/3544548.3581518},
DOI={10.1145/3544548.3581518},
booktitle={Proceedings of the 2023 CHI Conference on Human Factors in Computing Systems},
publisher={ACM},
author={Bhat, Avinash and Coursey, Austin and Hu, Grace and Li, Sixian and Nahar, Nadia and Zhou, Shurui and Kästner, Christian and Guo, Jin L.C.},
year={2023},
month=apr,
pages={1–17},
collection={CHI ’23}
}

@article{Gebru_2021,
title={Datasheets for datasets},
volume={64},
ISSN={1557-7317},
url={http://dx.doi.org/10.1145/3458723},
DOI={10.1145/3458723},
number={12},
journal={Communications of the ACM},
publisher={Association for Computing Machinery (ACM)},
author={Gebru, Timnit and Morgenstern, Jamie and Vecchione, Briana and Vaughan, Jennifer Wortman and Wallach, Hanna and III, Hal Daumé and Crawford, Kate},
year={2021},
month=nov,
pages={86–92}
}

@inproceedings{Mitchell_2019,
series={FAT* ’19},
title={Model Cards for Model Reporting},
url={http://dx.doi.org/10.1145/3287560.3287596},
DOI={10.1145/3287560.3287596},
booktitle={Proceedings of the Conference on Fairness, Accountability, and Transparency},
publisher={ACM},
author={Mitchell, Margaret and Wu, Simone and Zaldivar, Andrew and Barnes, Parker and Vasserman, Lucy and Hutchinson, Ben and Spitzer, Elena and Raji, Inioluwa Deborah and Gebru, Timnit},
year={2019},
month=jan,
pages={220–229},
collection={FAT* ’19}
}

@article{Retzlaff_2024,
title={Post-hoc vs ante-hoc explanations: xAI design guidelines for data scientists},
volume={86},
ISSN={1389-0417},
url={http://dx.doi.org/10.1016/j.cogsys.2024.101243},
DOI={10.1016/j.cogsys.2024.101243},
journal={Cognitive Systems Research},
publisher={Elsevier BV},
author={Retzlaff, Carl O. and Angerschmid, Alessa and Saranti, Anna and Schneeberger, David and Röttger, Richard and Müller, Heimo and Holzinger, Andreas},
year={2024},
month=aug,
pages={101243}
}

@article{Kalinowski_2025,
title={Naming the Pain in machine learning-enabled systems engineering},
volume={187},
ISSN={0950-5849},
url={http://dx.doi.org/10.1016/j.infsof.2025.107866},
DOI={10.1016/j.infsof.2025.107866},
journal={Information and Software Technology},
publisher={Elsevier BV},
author={Kalinowski, Marcos and Mendez, Daniel and Giray, Görkem and Santos Alves, Antonio Pedro and Azevedo, Kelly and Escovedo, Tatiana and Villamizar, Hugo and Lopes, Helio and Baldassarre, Teresa and Wagner, Stefan and Biffl, Stefan and Musil, Jürgen and Felderer, Michael and Lavesson, Niklas and Gorschek, Tony},
year={2025},
month=nov,
pages={107866}
}

@inbook{Haug_2025,
title={MLOps Adoption in the Manufacturing Industry: A Case Study with Zeiss SMT},
ISBN={9783032073136},
ISSN={1865-0937},
url={http://dx.doi.org/10.1007/978-3-032-07313-6_2},
DOI={10.1007/978-3-032-07313-6_2},
booktitle={Service-Oriented Computing},
publisher={Springer Nature Switzerland},
author={Haug, Markus and Azamnouri, Aidin and Fritz, Manuel and Woltmann, Lucas and Schriever, Christian and Wagner, Stefan},
year={2025},
month=oct,
pages={16–36}
}

@article{Latendresse_2024,
title={An Exploratory Study on Machine Learning Model Management},
volume={34},
ISSN={1557-7392},
url={http://dx.doi.org/10.1145/3688841},
DOI={10.1145/3688841},
number={1},
journal={ACM Transactions on Software Engineering and Methodology}, publisher={Association for Computing Machinery (ACM)},
author={Latendresse, Jasmine and Abedu, Samuel and Abdellatif, Ahmad and Shihab, Emad},
year={2024},
month=dec,
pages={1–31}
}

@inproceedings{Indykov_2025,
series={SAC ’25},
title={Quality trade-offs in ML-enabled systems: a multiple-case study}, url={http://dx.doi.org/10.1145/3672608.3707754},
DOI={10.1145/3672608.3707754},
booktitle={Proceedings of the 40th ACM/SIGAPP Symposium on Applied Computing},
publisher={ACM},
author={Indykov, Vladislav and Wohlrab, Rebekka and Strüber, Daniel}, year={2025},
month=mar,
pages={1730–1737},
collection={SAC ’25}
}

@article{Assres_2025,
title={State-of-the-Art and Challenges of Engineering ML- Enabled Software Systems in the Deep Learning Era},
volume={57},
ISSN={1557-7341},
url={http://dx.doi.org/10.1145/3731597},
DOI={10.1145/3731597},
number={10},
journal={ACM Computing Surveys},
publisher={Association for Computing Machinery (ACM)},
author={Assres, Gebremariam and Bhandari, Guru and Shalaginov, Andrii and Gronli, Tor-Morten and Ghinea, Gheorghita},
year={2025},
month=may,
pages={1–35}
}

@book{Guest_2012,
title={Applied Thematic Analysis},
ISBN={9781483384436},
url={http://dx.doi.org/10.4135/9781483384436},
DOI={10.4135/9781483384436},
publisher={SAGE Publications, Inc.},
author={Guest, Greg and MacQueen, Kathleen and Namey, Emily},
year={2012}
}

@book{Saldana_2021,
title={The Coding Manual for Qualitative Researchers},
ISBN={9781036235611},
url={http://dx.doi.org/10.4135/9781036235611},
DOI={10.4135/9781036235611},
publisher={SAGE Publications Ltd},
author={Saldana, Johnny},
year={2021}
}

\end{document}